\documentclass{article}
\usepackage{amsmath}
\usepackage{graphicx}

\newcommand{\ket}[1]{|{#1}\rangle}

\begin{document}
\title{Quantum interference in optical fields and atomic radiation}
\author{Z. Ficek$^{1}$ and S. Swain$^{2}$\\
$^{2}$Department of Physics and Centre for Laser Science, \\
The University of Queensland, Brisbane, QLD 4072, Australia \\
[8pt]\\
$^{2}$Department of Applied Mathematics and Theoretical Physics,\\
The Queen's University of Belfast, Belfast BT7 1NN, Northern Ireland}
\date{\today}

\maketitle

\begin{abstract}
We discuss the connection between quantum interference effects in optical
beams and radiation fields emitted from atomic systems. We illustrate this
connection by a study of the first- and second-order correlation functions
of optical fields and atomic dipole moments. We explore the role of
correlations between the emitting systems and present examples of practical
methods to implement two systems with non-orthogonal dipole moments. We also
derive general conditions for quantum interference in a two-atom system and
for a control of spontaneous emission. The relation between population
trapping and dark states is also discussed. Moreover, we present quantum
dressed-atom models of cancellation of spontaneous emission, amplification
on dark transitions, fluorescence quenching and coherent population trapping.
\end{abstract}

\section{Introduction}

\label{sec1}

Optical interference is a very old technique which began with Michelson's
and Young's experiments~\cite{mw}. In these experiments, a beam of light is
divided into two beams and, after traveling a distance long compared to the
optical wavelength, these two beams are recombined at an observation point.
If there is a small path difference between the beams, interference fringes
are found at the observation point. The interference fringes are a
manifestation of temporal coherence (Michelson interferometer) or spatial
coherence (Young interferometer) between the two light beams. The
interference experiments played a central role in early discussions of the
dual nature of light, and the appearance of an interference pattern was
recognized as a demonstration that light is a wave. The interpretation of
the interference experiments changed with the birth of quantum mechanics,
when corpuscular properties of light showed up in many experiments.
According to the quantum mechanical interpretation, given by Dirac~\cite
{dirac}, the interference pattern observed in the Young's experiment results
from a superposition of the probability amplitudes of a single photon to
take either of two possible pathways.

Interference effects can be observed not only between two light beams but
also between radiation fields emitted from a small number of atoms or even
from a single multilevel atom~\cite{szqo}. The interest in quantum
interference in atomic system stems from the early 1970s when Agarwal~\cite
{ag74} showed that the ordinary spontaneous decay of an excited degenerate $V
$-type three-level atom can be modified due to interference between the two
atomic transitions. The analysis of quantum interference has since been
extended to multiatom systems and multi-level atoms. Many interesting
effects have been predicted and a very wide range of practical applications
have been proposed. These are too numerous to detail here, but we may
mention some of the `traditional' applications, such as control of optical
properties of quantum systems, including high-contrast resonances~\cite
{zs96,ke}, electromagnetically induced transparency~\cite{bol}, lasing
without inversion \cite{mo-co00}, amplification without population inversion~
\cite{har}, enhancement of the index of refraction without absorption~\cite
{riwa}, and ultra-slow velocities of light~\cite{hau}.

In addition, quantum interference has been recognized as the most
significant mechanism for the modification and suppression of spontaneous
emission. The control and suppression of spontaneous emission, which is a
source of undesirable noise (decoherence), is very significant in the
context of quantum computation, teleportation, and quantum information
processing.

The effect of quantum interference on spontaneous emission in atomic and
molecular systems is the generation of superposition states which can be
manipulated by adjusting the polarizations of the transition dipole moments,
or the amplitudes  of the external driving fields. The spontaneous emission
can also be controlled by manipulating the lasers' phases \cite{phase}. With
a suitable choice of parameters, the superposition states can decay with
controlled and significantly reduced rates. This modification can lead to
subnatural linewidths in the fluorescence and absorption spectra~\cite
{zs96,zs97} and population trapping~\cite{dark,dicke}. Although the trapping
states have the common property that the population will stay in such a
state for an extremely long time, they can however be implemented in
different ways. In a multi-level system the population can be trapped in a
linear superposition of the bare atomic states, or in a dressed state
corresponding to an eigenstate of the atoms plus external fields, or in some
cases, in one of the excited states of the system.

In contrast to a simple theoretical picture of the effect of quantum
interference on spontaneous emission in atomic systems, experimental work
has proved to be extremely difficult, with only one experiment so far
demonstrating the constructive and destructive interference effects in
spontaneous emission~\cite{xia,li}. In the experiment they used sodium
dimers, which can be modeled as five-level molecular systems with a single
ground level, two intermediate and two upper levels, driven by a two-photon
process from the ground level to the upper doublet. By monitoring the
fluorescence from the upper levels they observed that the total fluorescent
intensity, as a function of two-photon detuning, is composed of two peaks on
transitions with parallel and three peaks on transitions with antiparallel
dipole moments. The observed variation of the number of peaks with the
mutual polarization of the dipole moments gives compelling evidence for
quantum interference in spontaneous emission. Agarwal~\cite{ag97} has
provided an intuitive picture for the observed spontaneous emission
cancellation in terms of interference pathways involving a two-photon
absorption process. Berman~\cite{ber} has shown that the experimentally
observed cancellation of spontaneous emission involving a two-photon
absorption process can be interpreted in terms of population trapping.
Although a cancellation of spontaneous emission is present with a two-photon
excitation process, no variation of the number of peaks with the
polarization of the dipole moments exist in the fluorescent intensity.
Recently, Wang {\it et al.}~\cite{wang} have presented a theoretical model
of the observed fluorescence intensity which explains the variation of the
number of the observed peaks with the mutual polarization of the molecular
dipole moments.

In this paper we discuss the connection between quantum interference with
optical beams and radiation fields emitted from atomic systems. We begin in
Section~\ref{sec2} by presenting basic concepts and definitions of the
first- and second-order correlation functions, which are frequently used in
the analysis of the interference phenomena. Section~\ref{sec3} describes the
master equation approach to quantum interference effects in atomic and
molecular systems. In Sections~\ref{sec4} we discuss different methods to
implement two transitions with non-orthogonal dipole moments. In Section~\ref
{sec5}, we derive general conditions for quantum interference in a two-atom
system and discuss the role of the interatomic interactions. Next, in
Section~\ref{sec6}, we discuss general conditions for a control of
spontaneous emission from two coupled systems and explain the relation
between population trapping and dark states. In Section~\ref{sec7}, we
present quantum dressed-atom models of cancellation of spontaneous emission,
amplification on dark transitions, fluorescence quenching, and coherent
population trapping.

\section{Optical interference and coherence}

\label{sec2}

The phenomenon of optical interference is usually described in completely
classical terms, in which optical fields are represented by classical plane
waves. The classical theory readily explains the presence of an interference
pattern in the first-order optical coherence. However, there are
higher-order interference effects that distinguish the quantum nature of
light from the wave nature.

\subsection{First-order interference}

\label{sec2a}

The simplest example for a demonstration of the first-order optical
interference is the Young's double slit experiment in which two light beams
of amplitudes ${\bf E}_{1}\left({\bf r}_{1}, t_{1}\right)$ and ${\bf E}%
_{2}\left({\bf r}_{2}, t_{2}\right)$, produced at two slits located at ${\bf %
r}_{1}$ and ${\bf r}_{2}$, respectively, incident on the screen at a point $P $, 
whose position vector is \textbf{R}. The resulting average intensity of the two fields detected at the point 
$P $ can be written as 

\begin{eqnarray}
\langle I\left({\bf R},t\right)\rangle &=& |u_{1}|^{2}\langle I_{1}\left(%
{\bf R}_{1},t-t_{1}\right)\rangle +|u_{2}|^{2}\langle I_{2}\left({\bf R}%
_{2},t-t_{2}\right)\rangle  \nonumber \\
&&+ 2\mbox{Re}\left\{u^{\ast}_{1}u_{2}G^{(1)}\left({\bf R}_{1},t-t_{1}; {\bf %
R}_{2},t-t_{2}\right)\right\} ,  \label{1}
\end{eqnarray}

where 
\begin{equation}
G^{(1)}\left({\bf R}_{1},t-t_{1};{\bf R}_{2},t-t_{2}\right) = \langle {\bf E}%
_{1}^{\ast}\left({\bf R}_{1}, t-t_{1}\right)\cdot {\bf E}_{2} \left({\bf R}%
_{2}, t-t_{2}\right)\rangle  \label{2}
\end{equation}
is the first order correlation function (coherence) between the field at $%
{\bf R}_{2}= \bf R -\bf r_{2} $ and the complex-conjugate field at 
${\bf R}_{1}= \bf R -\bf r_{1} $, at times $%
t-t_{2}$ and $t-t_{1}$, respectively, and $u_{i}$ is a constant which depends on
the geometry and the size of the $i$th slit. Here $ t_{i} = R_{i}/c $.

It is convenient to introduce the normalized first-order correlation
function as 
\begin{eqnarray}
g^{\left( 1\right) }\left( {\bf R}_{1},t_{1};{\bf R}_{2},t_{2}\right) &=&%
\frac{G^{\left( 1\right) }\left( {\bf R}_{1},t_{1};{\bf R}_{2},t_{2}\right) 
}{\sqrt{G^{\left( 1\right) }\left( {\bf R}_{1},t_{1};{\bf R}%
_{1},t_{1}\right) G^{\left( 1\right) }\left( {\bf R}_{2},t_{2};{\bf R}%
_{2},t_{2}\right) }}  \nonumber \\
&=&\frac{G^{\left( 1\right) }\left( {\bf R}_{1},t_{1};{\bf R}%
_{2},t_{2}\right) }{\sqrt{I_{1}\left( {\bf R}_{1},t_{1}\right) I_{2}\left( 
{\bf R}_{2},t_{2}\right) }},  \label{3}
\end{eqnarray}
satisfying the condition $0\leq |g^{\left( 1\right) }|\leq 1$. The
normalized correlation function~(\ref{3}) is often called the degree of
coherence and $g^{\left( 1\right) }=0$ for two independent fields, whereas $%
g^{\left( 1\right) }=1$ for perfectly correlated fields. The
intermediate values of $g^{\left( 1\right) }$ $(0<|g^{\left( 1\right) }|<1)$
characterize a partial coherence between the fields.

The average intensity $\langle I\left( {\bf R},t\right) \rangle $ depends on 
$g^{\left( 1\right) }$ and in the case of identical slits $(u_{1}=u_{2})$
and perfectly correlated fields $(|g^{\left( 1\right) }|=1)$, the
intensity can vary from $(\sqrt{I_{1}}-\sqrt{I_{2}})^{2}$ to $(\sqrt{I_{1}}+%
\sqrt{I_{2}})^{2}$, giving the so-called interference pattern. When $%
I_{1}=I_{2}=I_{0}$, the total average intensity varies from $\langle
I\rangle _{min}=0$ to $\langle I\rangle _{max}=4\langle I_{0}\rangle $,
giving maximal interference. For two independent fields, $g^{\left(
1\right) }=0$, and then the resulting intensity is just the sum of the
intensities of the two fields, that does not vary with the position of~$P$.

The usual measure of the depth of modulation (sharpness) of interference
fringes is the visibility in an interference pattern defined as 
\begin{equation}
{\cal {V}}\left({\bf R}\right) = \frac{\langle I\left({\bf R}%
,t\right)\rangle_{max} -\langle I\left({\bf R},t\right) \rangle_{min}}{%
\langle I\left({\bf R},t\right)\rangle_{max} +\langle I\left({\bf R}%
,t\right)\rangle_{min}} ,  \label{4}
\end{equation}
where $\langle I\left({\bf R},t\right)\rangle_{max}$ and $\langle I\left(%
{\bf R},t\right)\rangle_{min}$ represent the intensity maxima and minima at
the point $P$.

Since 
\begin{equation}
\langle I\rangle _{max}=\langle I_{1}\rangle +\langle I_{2}\rangle +2\sqrt{%
I_{1}I_{2}}|g^{(1)}|,  \label{5}
\end{equation}
and 
\begin{equation}
\langle I\rangle _{min}=\langle I_{1}\rangle +\langle I_{2}\rangle -2\sqrt{%
I_{1}I_{2}}|g^{(1)}|,  \label{6}
\end{equation}
we obtain 
\begin{equation}
{\cal {V}}\left( {\bf R}\right) =\frac{2\sqrt{I_{1}I_{2}}}{\left(
I_{1}+I_{2}\right) }|g^{(1)}|.  \label{7}
\end{equation}
Thus, $|g^{(1)}|$ determines the visibility of the interference fringes. In
the special case of equal intensities of the two fields $(I_{1}=I_{2})$, the
visibility simplifies to ${\cal {V}}\left( {\bf R}\right) =|g^{(1)}|$, i.e. $%
|g^{(1)}|$ is then equal to the visibility. For perfectly correlated fields $%
{\cal {V}}\left( {\bf R}\right) =1$, while ${\cal {V}}\left( {\bf R}\right)
=0$ for uncorrelated fields. When $I_{1}\neq I_{2}$, the visibility is
always smaller than one, even for perfectly correlated fields. This fact is
related to the problem of extracting which way information has been
transferred through the slits into the point $P$ ({\it welcher Weg }or{\it \
which-way} information). The observation of an interference pattern and the
acquisition of which-way information are mutually exclusive: a which-way
measurement necessarily destroys the interference fringes.

We can introduce an inequality according to which the fringe visibility $%
{\cal {V}}$ displayed at the point~$P$ and an absolute upper bound on the
amount of which-way information ${\cal {D}}$ that can be detected at the
point $P,$ defined as the difference in the probabilities for taking either
of the two paths, are related by~\cite{eng} 
\begin{equation}
{\cal {D}}^{2}+{\cal {V}}^{2}\leq 1.  \label{8}
\end{equation}
Hence, the extreme situations characterized by perfect fringe visibility $(%
{\cal {V}}=1)$ or full knowledge of which-way information has been
transmitted $({\cal {D}}=1)$ are mutually exclusive. In order to distinguish
which-way information has been transmitted, one can locate an intensity
detector at the point $P$ and adjust it to measure a field of a particular
intensity $I_{d}$. When the fields coming from the slits have the same
intensities, the detector cannot distinguish which-way the fields came to
the point $P$, so there is no which-way information available $({\cal {D}}=0)
$ resulting in perfect fringe visibility $({\cal {V}}=1)$. On the other
hand, when the intensities of the fields are different $(I_{1}\neq I_{2})$,
the detector adjusted to measure a particular intensity can distinguish
which way the field came to the point $P$ resulting in the disappearance of
the interference fringes. This is clearly seen from Eq.~(\ref{7}): if $%
I_{1}\gg I_{2}$ or $I_{1}\ll I_{2}$, the visibility ${\cal {V}}\approx 0$
even for $|g^{(1)}|=1$. The same arguments apply to the frequencies and
phases of the detected fields.

Information about the frequencies and phases of the detected fields is
provided by the argument and phase of $g^{(1)}$. Moreover, the phase of $%
g^{(1)}$ determines the positions of the fringes in the interference
pattern. If the observation point $P$ lies in the far field zone of the
radiation emitted by the slits, the fields at the observation point can be
approximated by plane waves which may be written as 
\begin{equation}
{\bf E}\left( {\bf R}_{i},t-t_{i}\right) \approx {\bf E}\left( {\bf R}%
_{i},t\right) \exp \left[ -i\left( \omega _{i}t_{i}+\phi _{i}\right) \right]
={\bf E}\left( {\bf R}_{i},t\right) \exp \left[ -i\left( \omega
_{i}R_{i}/c+\phi _{i}\right) \right] ,\quad i=1,2,  \label{9}
\end{equation}
where $\omega _{i}$ is the angular frequency of the $i$th field and $\phi
_{i}$ is its initial phase. We can measure the frequencies from the average
frequency of the two fields as 
\begin{eqnarray}
\omega _{1} &=&\omega _{0}+\frac{1}{2}\Delta ,  \nonumber \\
\omega _{2} &=&\omega _{0}-\frac{1}{2}\Delta ,  \label{10}
\end{eqnarray}
where $\omega _{0}=(\omega _{1}+\omega _{2})/2$ is the average frequency, 
and $\Delta =\omega _{1}-\omega _{2}$.

Since the observation point lies in the far field zone of the radiation emitted
by the slits, i.e. the separation between the slits is very small compared
to the distance to the point $P$, we can write approximately 
\begin{equation}
R_{i}=|{\bf R}-{\bf r}_{i}|\approx R-\bar{{\bf R}}\cdot {\bf r}_{i},
\label{11}
\end{equation}
where $\bar{{\bf R}}={\bf R}/R$ is the unit vector in the direction ${\bf R}$%
. Hence, substituting Eq.~(\ref{9}) with (\ref{10}) and (\ref{11}) into Eq.~(%
\ref{3}), we obtain 
\begin{equation}
g^{\left( 1\right) }\left( {\bf R}_{1},t_{1};{\bf R}_{2},t_{2}\right)
=|g^{\left( 1\right) }\left( {\bf R}_{1},t;{\bf R}_{2},t\right) |\exp \left(
ik_{0}\bar{{\bf R}}\cdot {\bf r}_{21}\right) \exp \left[ i\left( k_{0}\tilde{%
R}\frac{\Delta }{\omega _{0}}+\delta \phi \right) \right] ,  \label{12}
\end{equation}
where ${\bf r}_{21}={\bf r}_{2}-{\bf r}_{1}$ is the distance between the
slits, $\tilde{R}=R+\frac{1}{2}\bar{{\bf R}}\cdot \left( {\bf r}_{1}+{\bf r}%
_{2}\right) $, $\delta \phi =\phi _{1}-\phi _{2}$, $k_{0}=\omega _{0}/c=2\pi
/\lambda _{0}$, and $\lambda _{0}$ represents the mean wavelength of the
fields. Let us analyze the physical meaning of the exponents appearing on
the right-hand side of Eq.~(\ref{12}). The first exponent depends on the
separation between the slits and the position~${\bf R}$ of the point~$P$.
For small separations the exponent changes slowly with the position~${\bf R}$
and leads to minima and maxima in the interference pattern. The minima
appear whenever 
\begin{equation}
k_{0}\bar{{\bf R}}\cdot {\bf r}_{21}=\left( 2n+1\right) \pi ,\quad n=0,\pm
1,\pm 2,\ldots .  \label{13}
\end{equation}
The second exponent, appearing in Eq.~(\ref{12}), depends on the sum of the
position of the slits, the ratio $\Delta /\omega _{0}$ and the difference $%
\delta \phi $ between the initial phases of the fields. This term introduces
limits on the visibility of the interference pattern and can affect the
pattern only if the frequencies and the initial phases of the fields are
different. Even for equal and well stabilized phases, but significantly
different frequencies of the fields such that $\Delta /\omega _{0}\approx 1$%
, the exponent oscillates rapidly with ${\bf R}$ leading to the
disappearance of the interference pattern. Thus, in order to observe an
interference pattern it is important to have two fields of well stabilized
phases and equal or nearly equal frequencies. Otherwise, no interference
pattern can be observed even if the fields are perfectly correlated.

The dependence of the interference pattern on the frequencies and phases of
the fields is related to the problem of extracting which-way information.
Consider the case where the two fields have mean intensities: $%
I_{1}=I_{2}=I_{0}.$ For perfectly correlated fields with equal frequencies $%
(\Delta =0)$ and equal initial phases $(\phi _{1}=\phi _{2})$, the total
intensity at the point~$P$ is 
\begin{equation}
\langle I\left( {\bf R}\right) \rangle =2\langle I_{0}\rangle \left( 1+\cos
k_{0}\bar{{\bf R}}\cdot {\bf r}_{21}\right) ,  \label{14}
\end{equation}
giving an interference pattern with the maximum visibility of
100\%. There are no features of the two beams which can be used to
distinguish between them. For general frequencies and phases, the total
intensity at the point $P$ is given by 
\begin{eqnarray}
\langle I\left( {\bf R}\right) \rangle &=&2\langle I_{0}\rangle \left[
1+\left( \cos k_{0}\bar{{\bf R}}\cdot {\bf r}_{21}\right) \cos \left( k_{0}%
\tilde{R}\frac{\Delta }{\omega _{0}}+\delta \phi \right) \right.  \nonumber
\\
&&-\left. \left( \sin k_{0}\bar{{\bf R}}\cdot {\bf r}_{21}\right) \sin
\left( k_{0}\tilde{R}\frac{\Delta }{\omega _{0}}+\delta \phi \right) \right]
.  \label{15}
\end{eqnarray}
In this case the intensity at $P$ exhibits additional cosine and sine
modulations, and at the minima the intensity is non-zero indicating that
100\% modulation is not possible for two fields of different frequencies
and/or initial phases. Moreover, for large frequency differences $(\Delta
/\omega _{0}\gg 1)$ the cosine and sine terms oscillate rapidly with ${\bf R}
$ and average to zero, washing out the interference pattern. In terms of
which-way information, a detector adjusted to measure a particular frequency
or phase could distinguish between the two fields. Clearly, one could tell
which way the detected field came to the point $P$.

Thus, whether which-way information is available or not depends on the
frequencies and phases as well as the intensities of the interfering fields.
Maximum possible which-way information results in destruction of the
interference pattern, and vice versa, a complete lack of which-way
information results in maximum visibility of the interference pattern.

\subsection{Second-order interference}

\label{sec2b}

The analysis of the interference phenomenon can be extended into
higher-order correlation functions. The first experimental demonstration
that such correlations exist in optical fields was given by Hanbury-Brown
and Twiss~\cite{hbt}, who measured the second-order correlation function of
a thermal field.

The second-order (intensity) correlation function of a field of a complex
amplitude ${\bf E}\left({\bf R},t\right)$ is defined as 
\begin{eqnarray}
G^{(2)}\left({\bf R}_{1},t_{1}; {\bf R}_{2},t_{2}\right) &=& \langle {\bf E}%
^{\ast}\left({\bf R}_{1},t_{1}\right){\bf E}^{\ast} \left({\bf R}%
_{2},t_{2}\right){\bf E}\left({\bf R}_{2}, t_{2}\right) {\bf E}\left({\bf R}%
_{1}, t_{1}\right)\rangle  \nonumber \\
&=& \langle I\left({\bf R}_{1},t_{1}\right)I\left({\bf R}_{2},t_{2}\right)
\rangle ,  \label{16}
\end{eqnarray}
where $I\left({\bf R}_{1},t_{1}\right)$ and $I\left({\bf R}_{2},t_{2}\right)$
are the instantaneous intensities of the field detected at a point ${\bf R}%
_{1}$ at time $t_{1}$ and at a point ${\bf R}_{2}$ at time $t_{2}$,
respectively.

The second-order correlation function has completely different coherence
properties than the first-order correlation function. An interference
pattern can be observed in the second-order correlation function, but in
contrast to the first-order correlation function, the interference appears
between two points located at ${\bf R}_{1}$ and ${\bf R}_{2}$. Moreover, an
interference pattern can be observed even if the fields are produced by two
independent sources for which the phase difference $\phi _{1}-\phi _{2}$ is
completely random~\cite{man65}. In this case the second-order correlation
function is given by 
\begin{equation}
G^{(2)}\left( {\bf R}_{1},t_{1};{\bf R}_{2},t_{2}\right) =\langle
I_{1}^{2}\left( t_{1}\right) \rangle +\langle I_{2}^{2}\left( t_{2}\right)
\rangle +2\langle I_{1}\left( t_{1}\right) I_{2}\left( t_{2}\right) \rangle %
\left[ 1+\cos k{\bf r}_{21}\cdot \left( \bar{{\bf R}}_{1}-\bar{{\bf R}}%
_{2}\right) \right] .  \label{17}
\end{equation}
Clearly, the second-order correlation function of two independent fields
exhibits a cosine modulation with the separation ${\bf R}_{1}-{\bf R}_{2}$
of the two detectors. This is an interference effect although it involves a
correlation function that is second order in the intensity. Similarly to the
first-order correlation function, the sharpness of the fringes depends on
the relative intensities of the fields. For equal intensities, $%
I_{1}=I_{2}=I_{0}$, the correlation function~(\ref{17}) reduces to 
\begin{equation}
G^{(2)}\left( {\bf R}_{1},t;{\bf R}_{2},t\right) =4\langle I_{0}^{2}\rangle %
\left[ 1+\frac{1}{2}\cos k{\bf r}_{21}\cdot \left( \bar{{\bf R}}_{1}-\bar{%
{\bf R}}_{2}\right) \right] .  \label{18}
\end{equation}

In analogy to the visibility in the first-order correlation function, we can
define the visibility of the interference pattern of the intensity
correlations as 
\begin{equation}
{\cal {V}}=\frac{G_{max}^{(2)}-G_{min}^{(2)}}{G_{max}^{(2)}+G_{min}^{(2)}},
\label{19}
\end{equation}
and find from~Eq.~(\ref{18}) that in the case of classical fields the
maximum possible visibility of the interference pattern that can be observed
is ${\cal {V}}=\frac{1}{2}$. That is, two independent fields with random and
uncorrelated phases can exhibit an interference pattern in the intensity
correlation with a maximum visibility of $50\%$.

\subsection{Quantum interference}

\label{sec2c}

In the classical theory of light and coherence the field is represented by
complex vectorial amplitudes ${\bf E}\left( {\bf r},t\right) $ and ${\bf E}%
^{\ast }\left( {\bf r},t\right) $, which are complex numbers (c-numbers). In
the quantum theory of light the most important physical quantity is the
electric field, which is represented by the field {\em operator} $\hat{{\bf E%
}}\left( {\bf r},t\right) $. The coherence properties are discussed in terms
of the negative and positive frequency parts $\hat{{\bf E}}^{(+)}$ and $\hat{%
{\bf E}}^{(-)}$ of the total field operator $\hat{{\bf E}}$ \cite{glaub}.

In the quantum description of the field, the first- and second-order
correlation functions are defined in terms of the normally-ordered field
operators $\hat{{\bf E}}^{(+)}$ and $\hat{{\bf E}}^{(-)}$ as 
\begin{eqnarray}
G^{(1)}\left( {\bf R}_{1},t_{1};{\bf R}_{2},t_{2}\right) &=&\langle \hat{%
{\bf E}}^{(-)}\left( {\bf R}_{1},t_{1}\right) \cdot \hat{{\bf E}}%
^{(+)}\left( {\bf R}_{2},t_{2}\right) \rangle ,  \nonumber \\
G^{(2)}\left( {\bf R}_{1},t_{1};{\bf R}_{2},t_{2}\right) &=&\langle \hat{%
{\bf E}}^{(-)}\left( {\bf R}_{1},t_{1}\right) \hat{{\bf E}}^{(-)}\left( {\bf %
R}_{2},t_{2}\right) \hat{{\bf E}}^{(+)}\left( {\bf R}_{2},t_{2}\right) \hat{%
{\bf E}}^{(+)}\left( {\bf R}_{1},t_{1}\right) \rangle ,  \label{20}
\end{eqnarray}
where the average is taken over an initial state of the field \cite{glaub}.

The correlation functions~(\ref{20}) described by the field operators are
formally similar to the correlation functions~(\ref{2}) and~(\ref{16}) of
the classical field. A closer examination of Eqs.~(\ref{2}), (\ref{16}),
and~(\ref{20}) shows that the first-order correlation functions do not
distinguish between the quantum and classical theories of the
electromagnetic field. However, there are significant differences between
the classical and quantum descriptions of the field in the properties of the
second-order correlation function~\cite{mr}.

As an example, consider the simple case of two single-mode fields of equal
frequencies and polarizations. Assume that there are initially $n$ photons
in the mode~$a$ and $m$ photons in the mode $b$, and the state vectors of
the fields are the Fock states $|\psi _{a}\rangle =|n\rangle $ and $|\psi
_{b}\rangle =|m\rangle $. The initial state of the two fields is the direct
product of the single-field states, $|\psi \rangle =|n\rangle |m\rangle $.
Inserting Eq.~(\ref{19}) into Eq.~(\ref{20}) and taking the expectation
value with respect to the initial state of the fields, we find 
\begin{equation}
G^{(2)}\left( {\bf R}_{1},t_{1};{\bf R}_{2},t_{2}\right) =\left( \frac{\hbar
\omega }{2\epsilon _{0}V}\right) ^{2}\left\{ n\left( n-1\right) +m\left(
m-1\right) +2nm\left[ 1+\cos k{\bf r}_{21}\cdot \left( \bar{{\bf R}}_{1}-%
\bar{{\bf R}}_{2}\right) \right] \right\} .  \label{21}
\end{equation}
It is seen that the first two terms on the right-hand side of Eq.~(\ref{21})
vanish when the number of photons in each mode is smaller than 2, i.e. $n<2$
and $m<2$. In this limit the correlation function~(\ref{21}) reduces to 
\begin{equation}
G^{(2)}\left( {\bf R}_{1},t_{1};{\bf R}_{2},t_{2}\right) =2\left( \frac{%
\hbar \omega }{2\epsilon _{0}V}\right) ^{2}\left[ 1+\cos k{\bf r}_{21}\cdot
\left( \bar{{\bf R}}_{1}-\bar{{\bf R}}_{2}\right) \right] .  \label{22}
\end{equation}
Thus, perfect interference pattern with the visibility ${\cal {V}}=1$ can be
observed in the second-order correlation function of two quantum fields each
containing only one photon. As we have noted, the classical theory predicts
a maximum visibility of ${\cal {V}}=0.5$. For $n,m\gg 1$, $m(m-1)\approx
n(n-1)\approx n^{2}$, and then the quantum correlation function (\ref{21})
reduces to that of the classical field.

It follows from Eq.~(\ref{22}) that the second-order correlation function of
the quantum field vanishes when 
\begin{equation}
k{\bf r}_{21}\cdot \left( \bar{{\bf R}}_{1}-\bar{{\bf R}}_{2}\right) =\left(
2n+1\right) \pi ,\quad n=0,\pm 1,\pm 2\ldots  \label{23}
\end{equation}
In other words, two photons can never be detected at two points separated by
an odd multiple of $\lambda r_{12} /2$, despite the fact that one photon can be
detected anywhere. The vanishing of the second-order correlation function
for two photons at widely separated points ${\bf R}_{1}$ and ${\bf R}_{2}$
is an example of quantum-mechanical nonlocality, that is the outcome of a
detection measurement at ${\bf R}_{1}$ appears to be influenced by where we
have chosen to locate the ${\bf R}_{2}$ detector. At certain positions ${\bf %
R}_{2}$ we can never detect a photon at ${\bf R}_{1}$ when there is a photon
detected at ${\bf R}_{2}$, whereas at other position ${\bf R}_{2}$ it is
possible. The photon correlation argument shows clearly that quantum theory
does not in general describe an objective physical reality independent of
observation~\cite{walls}.

The visibility of the fringes of the intensity correlations provides a means
of testing for quantum correlations between two light fields. Mandel~{\it et
al.}~\cite{mand} have measured the visibility in the interference of signal
and idler modes simultaneously generated in the process of degenerate
parametric down conversion, and observed a visibility of about $75\%$, that
is a clear violation of the upper bound of $50\%$ allowed by classical
correlations. Richter~\cite{rich} have extended the analysis of the
visibility into the third-order correlation function, and have also found
significant differences in the visibility of the interference pattern of the
classical and quantum fields.

\section{Quantum interference in atomic systems}

\label{sec3}

The phenomenon of optical interference can be observed not only between two
light beams but also between radiation fields emitted from a small number of
atoms or even in the radiation field emitted from a single multilevel
(multi-channel) system~\cite{szqo,ag74}. The atoms or atomic transitions can
be regarded as point sources of radiation, similar to the slits in Young's
original experiment. In this case, interference results from a superposition
of the transition amplitudes between quantum states of the atoms, and this
phenomenon has been designated as quantum interference. The essential
feature of quantum interference is the existence of linear superpositions of
the atomic states which can be induced by external or internal fields, or
even by the coupling of the atomic states through the environment (the
vacuum field).

\subsection{Correlation functions for atomic operators}

\label{sec3a}

Where atoms act as a source of the EM field, the correlation functions of
the emitted field can be related to the correlation functions of the atomic
variables, such as the atomic dipole operators.

The relation between the positive frequency part of the electric field
operator at a point ${\bf R}=R\bar{{\bf R}}$, in the far-field zone, and the
atomic dipole moments is given by the well-known expression~\cite{ag74,lem} 
\begin{equation}
\hat{{\bf E}}^{(+)}\left( {\bf R},t\right) =\hat{{\bf E}}_{0}^{(+)}\left( 
{\bf R},t\right) -\frac{1}{c^{2}}\sum_{i=1}^{2}\frac{\bar{{\bf R}}\times
\left( \bar{{\bf R}}\times \mbox{\boldmath $\mu$}_{i}\right) }{R}\omega
_{i}S_{i}^{-}\left( t-\frac{R}{c}\right) \exp \left[ -ik\bar{{\bf R}}\cdot 
{\bf r}_{i}\right] ,  \label{24}
\end{equation}
where $S_{i}^{-}$ is the dipole lowering operator of the $i$th atom (dipole
transition), $\mbox{\boldmath $\mu$}_{i}$ and $\omega _{i}$ are the
transition dipole matrix element and the angular frequency respectively, $%
{\bf r}_{i}$ is the position of the $i$th atom, and $\hat{{\bf E}}%
_{0}^{(+)}\left( {\bf R},t\right) $ denotes the positive frequency part of
the field in the absence of the atoms.

If we assume that initially the field is in the vacuum state, then the
free-field part $\hat{{\bf E}}_{0}^{(+)}\left( {\bf R},t\right) $ does not
contribute to the expectation values of the normally-ordered operators.
Hence, substituting Eq.(\ref{24}) into Eq.(\ref{20}) and integrating over
all directions ($4\pi $ solid angle) of spontaneous emission, we obtain the
following expressions for the first- and second-order correlation functions 
\begin{equation}
G^{\left( 1\right) }\left( {\bf R},t\right) =\sum_{i,j=1}^{2}\Gamma
_{ij}\left\langle S_{i}^{+}\left( t\right) S_{j}^{-}\left( t\right)
\right\rangle ,  \label{25}
\end{equation}
and 
\begin{equation}
G^{\left( 2\right) }\left( {\bf R},t_{1};{\bf R},t_{2}\right)
=\sum_{i,j,k,l=1}^{2}\Gamma _{il}\Gamma _{jk}\left\langle S_{i}^{+}\left(
t_{1}\right) S_{j}^{+}\left( t_{2}\right) S_{k}^{-}\left( t_{2}\right)
S_{l}^{-}\left( t_{1}\right) \right\rangle ,  \label{26}
\end{equation}
where $\Gamma _{ii}=\Gamma _{i}$ is the spontaneous decay rate of the $i$th
transition, while $\Gamma _{ij}$ is the so-called cross-damping rate arising
from the vacuum induced coupling between the two dipole moments.

For two dipole moments in a single atom, the cross-damping rate is given by 
\begin{equation}
\Gamma _{ij}=\frac{2\sqrt{\omega _{i}^{3}\omega _{j}^{3}}}{3\hbar c^{3}} %
\mbox{\boldmath $\mu$}_{i}\cdot \mbox{\boldmath $\mu$}_{j} = \sqrt{\Gamma
_{i}\Gamma _{j}}\cos \theta ,\qquad \left( i\neq j\right) ,  \label{27}
\end{equation}
where $\theta$ is the angle between the dipole moments. The cross-damping
rate is sensitive to the mutual polarization of the dipole moments. If the
dipole moments are parallel, $\theta =0^{o}$, and the cross-damping rate is
maximal with $\Gamma _{12}=\sqrt{\Gamma _{1} \Gamma _{2}}$, whilst $\Gamma
_{12}=0$ if the dipole moments are perpendicular $(\theta =90^{o})$.

In the case of two dipole transitions in two separate atoms, the
cross-damping rate depends not only on the orientation of the dipole moments
but also on the separation between the atoms, and is given by~\cite{ag74,lem}
\begin{eqnarray}
\Gamma _{12} &=&\frac{3}{4}\sqrt{\Gamma _{1}\Gamma _{2}}\left\{ \left[
1-\left( \bar{\mbox{\boldmath $\mu$}}\cdot \bar{{\bf r}}_{12}\right) ^{2}%
\right] \frac{\sin \left( k_{0}r_{12}\right) }{k_{0}r_{12}}\right.  \nonumber
\\
&&\left. +\left[ 1-3\left( \bar{\mbox{\boldmath $\mu$}}\cdot \bar{{\bf r}}%
_{12}\right) ^{2}\right] \left[ \frac{\cos \left( k_{0}r_{12}\right) }{%
\left( k_{0}r_{12}\right) ^{2}}-\frac{\sin \left( k_{0}r_{12}\right) }{%
\left( k_{0}r_{12}\right) ^{3}}\right] \right\} ,  \label{28}
\end{eqnarray}
where $\bar{\mbox{\boldmath $\mu$}}$ is the unit vector along the dipole
moments of the atoms, which we have assumed to be parallel $(\bar{%
\mbox{\boldmath $\mu$}}=\bar{\mbox{\boldmath $\mu$}}_{1}=\bar{%
\mbox{\boldmath $\mu$}}_{2})$, and $\bar{{\bf r}}_{12}$ is the unit vector
along the interatomic axis. The parameter (\ref{28}) is called the
collective damping rate.

\subsection{Master equation}

\label{sec3b}

There are a number of theoretical approaches that can be used to calculate
quantum interference effects in atomic systems. The traditional method is
the master equation approach in which the time evolution of the density
operator of radiating systems is given in terms of the damping rates $\Gamma
_{ij}$. For a system composed of two dipole transitions, the master equation
can be written as 
\begin{eqnarray}
\frac{\partial }{\partial t}\rho &=&-\frac{i}{\hbar }\left[
H_{s}+H_{sL}+H_{c},\rho \right]  \nonumber \\
&&-\frac{1}{2}\sum_{i,j=1}^{2}\Gamma _{ij}\left( S_{i}^{+}S_{j}^{-}\rho
+\rho S_{i}^{+}S_{j}^{-}-2S_{j}^{-}\rho S_{i}^{+}\right) ,  \label{29}
\end{eqnarray}
where 
\begin{equation}
H_{s}=\hbar \omega _{1}S_{1}^{+}S_{1}^{-}+\hbar \omega _{2}S_{2}^{+}S_{2}^{-}
\label{30}
\end{equation}
is the Hamiltonian of the systems, 
\begin{equation}
H_{sL}=-\frac{1}{2}\hbar \left\{ \Omega _{1}S_{1}^{+}\exp \left[ -i\left(
\omega _{L1}t+\phi _{1}\right) \right] +\Omega _{2}S_{2}^{+}\exp \left[
-i\left( \omega _{L2}t+\phi _{2}\right) \right] +{\rm H.c.}\right\}
\label{31}
\end{equation}
is the interaction of the atomic transitions with the coherent laser fields,
and 
\begin{equation}
H_{c}=\hbar \delta _{12}^{\left( -\right) }\left(
S_{1}^{+}S_{2}^{-}+S_{2}^{+}S_{1}^{-}\right) +\hbar \delta _{12}^{\left(
+\right) }\left( S_{1}^{-}S_{2}^{+}+S_{2}^{-}S_{1}^{+}\right) ,  \label{32}
\end{equation}
with 
\begin{equation}
\delta _{12}^{\left( \pm \right) }=\frac{P}{\pi }\int_{-\infty }^{\infty
}d\omega _{k}\frac{\Gamma _{12}}{\omega _{k}\pm \omega _{0}}  \label{33}
\end{equation}
is the frequency shift of the atomic transitions due to their mutual
interaction through the vacuum field.

In Eqs.~(\ref{29})-(\ref{33}), $\Omega _{1}$ and $\Omega _{2}$ are the Rabi
frequencies of the laser fields of angular frequencies $\omega _{L1},\omega
_{L2}$ and phases $\phi _{1},\phi _{2}$ respectively, $\omega _{0}=(\omega
_{1}+\omega _{2})/2$ and $P$ stands for the principal value of the integral.

The master equation gives us an elegant description of the physics involved
in the dynamics of two interacting atoms or atomic transitions. In the case
of two atoms, the cross-damping rate is given in Eq.~(\ref{28}) and the
frequency shifts $\delta _{12}^{(\pm )}$ are given by~\cite{ag74,lem} 
\begin{eqnarray}
\Omega _{12}=\delta _{12}^{(+)}+\delta _{12}^{(-)} &=&\frac{3}{4}\sqrt{%
\Gamma _{1}\Gamma _{2}}\left\{ -\left[ 1-\left( \bar{\mbox{\boldmath $\mu$}}%
\cdot \bar{{\bf r}}_{21}\right) ^{2}\right] \frac{\cos \left(
k_{0}r_{21}\right) }{k_{0}r_{21}}\right.  \nonumber \\
&&\left. +\left[ 1-3\left( \bar{\mbox{\boldmath $\mu$}}\cdot \bar{{\bf r}}%
_{21}\right) ^{2}\right] \left[ \frac{\sin \left( k_{0}r_{21}\right) }{%
\left( k_{0}r_{21}\right) ^{2}}+\frac{\cos \left( k_{0}r_{21}\right) }{%
\left( k_{0}r_{21}\right) ^{3}}\right] \right\} .  \label{34}
\end{eqnarray}
The parameter (\ref{34}) is called the retarded dipole-dipole interaction
between the atoms.

In the case of two coupled transitions in a single atom, the cross-damping
rate is given in Eq.~(\ref{27}) and $\delta_{12}^{(\pm)}$ are very small
shifts of the order of the Lamb shift~\cite{cs,afs}

The presence of the additional damping terms $\Gamma_{12}$ may suggest that
quantum interference enhances spontaneous emission from two coupled systems.
However, as we shall illustrate in the following sections, the cross-damping
rate can, in fact, lead to a reduction or even suppression of spontaneous
emission.

\section{Implementation of two transitions with non-orthogonal dipole moments%
}

\label{sec4}

Quantum interference between two transitions in a single atom may occur only
if the dipole moments of the transitions involved are non-orthogonal, i.e. 
\[
\mbox{\boldmath $\mu$}_{1}\cdot \mbox{\boldmath $\mu$}_{2}\neq 0. 
\]
This represents a formidable practical problem, as it is very unlikely to
find isolated atoms with two non-orthogonal dipole moments and quantum
states close in energy. Consider, for example, a $V$-type atom with the
upper states $|{1}\rangle $, $|{3}\rangle $ and the ground state $|{2}%
\rangle $. The evaluation of the dipole matrix elements produces the
following selection rules in terms of the angular momentum quantum numbers: $%
J_{1}-J_{2}=\pm 1,0$, $J_{3}-J_{2}=\pm 1,0$, and $M_{1}-M_{2}=M_{3}-M_{2}=%
\pm 1,0$. Since in many atomic systems $M_{1}\neq M_{3}$, then $%
\mbox{\boldmath $\mu$}_{12}$ is perpendicular to $\mbox{\boldmath $\mu$}%
_{32} $ and the atomic transitions are independent. Xia {\it et al.} have
found transitions with parallel and anti-parallel dipole moments in sodium
molecules (dimers) and have demonstrated experimentally the effect of
quantum interference on the fluorescence intensity, although their results
have been criticised~\cite{xia}. The transitions with parallel and
anti-parallel dipole moments in the sodium dimers result from a mixing of
the molecular states due to the spin-orbit coupling.

\subsection{External driving field method}

\label{sec4a}

A mixing of atomic or molecular states can be implemented by applying
external fields \cite{ha-ea91}. To illustrate this method, we consider a $V$%
-type atom with the upper states connected to the ground state by
perpendicular dipole moments $(\mbox{\boldmath $\mu$}_{12}\perp %
\mbox{\boldmath $\mu$}_{32})$. When the two upper states are coupled by a
resonant microwave field, the states become a linear superposition of the
bare states 
\begin{eqnarray}
|{a}\rangle &=&\frac{1}{\sqrt{2}}\left( |{1}\rangle +|{3}\rangle \right) , 
\nonumber \\
|{b}\rangle &=&\frac{1}{\sqrt{2}}\left( |{1}\rangle -|{3}\rangle \right) .
\label{36}
\end{eqnarray}
It is easy to find from Eq~(\ref{36}) that the dipole matrix elements
between the superposition states and the ground state $|{2}\rangle $ are 
\begin{eqnarray}
\mbox{\boldmath $\mu$}_{a2} &=&\frac{1}{\sqrt{2}}\left( 
\mbox{\boldmath
$\mu$}_{12}+\mbox{\boldmath $\mu$}_{32}\right) ,  \nonumber \\
\mbox{\boldmath $\mu$}_{b2} &=&\frac{1}{\sqrt{2}}\left( 
\mbox{\boldmath
$\mu$}_{12}-\mbox{\boldmath $\mu$}_{32}\right) .  \label{37}
\end{eqnarray}
When $|\mbox{\boldmath $\mu$}_{12}|\neq |\mbox{\boldmath $\mu$}_{32}|$, the
dipole moments $\mbox{\boldmath $\mu$}_{a2}$ and $\mbox{\boldmath $\mu$}%
_{b2} $ are not perpendicular. However, the dipole moments cannot be made
parallel or anti-parallel.

An alternative method in which one could create a $V$-type system with
parallel or anti-parallel dipole moments is to apply a strong laser field to
one of the two transitions in a $\Lambda $-type atom. The scheme is shown in
Fig.~1. When the dipole moments of the $|{1}\rangle \rightarrow |{3}\rangle $
and $|{2}\rangle \rightarrow |{3}\rangle $ transitions are perpendicular,
the laser exclusively couples to the $|{2}\rangle \rightarrow |{3}\rangle $
transition and produces dressed states 

\begin{figure}
\begin{center}
\includegraphics[height=2in,width=5in]{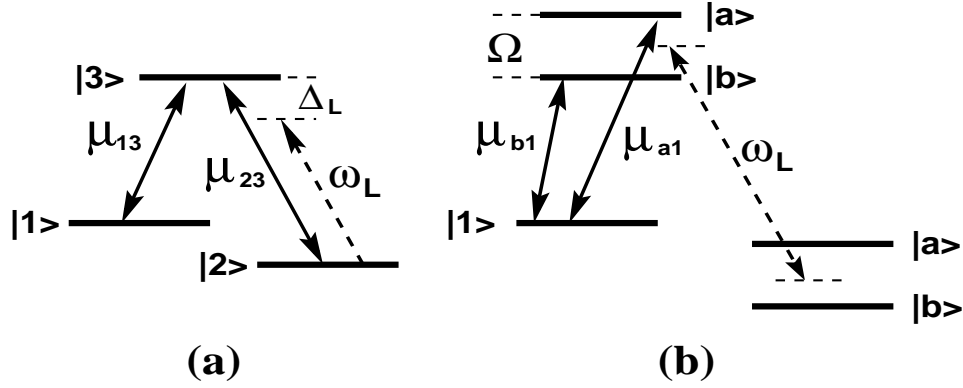}
\caption{Laser induced $V$-type system with non-degenerate transitions. A laser field applied to the $\ket 2 -\ket 3$ transition of a $\Lambda$ system creates non-degenerate dressed states separated by $\Omega =\sqrt{\Omega_{0}^{2}+\Delta^{2}_{L}}$. The sub-system with the upper dressed states $\ket a$, $\ket b$ and the ground state $\ket 1$ behaves as a $V$-type system with parallel dipole moments.}
\end{center}
\end{figure}

\begin{eqnarray}
|{a}\rangle  &=&\sin \phi |{2}\rangle +\cos \phi |{3}\rangle ,  \nonumber \\
|{b}\rangle  &=&\cos \phi |{2}\rangle -\sin \phi |{3}\rangle ,  \label{38}
\end{eqnarray}
where 
\begin{equation}
\cos ^{2}\phi =\frac{1}{2}+\frac{\Delta _{L}}{2\sqrt{\Omega _{0}^{2}+\Delta
_{L}^{2}}},  \label{39}
\end{equation}
$\Delta _{L}$ is the detuning of the laser frequency from the atomic
transition and $\Omega _{0}$ is the on-resonance Rabi frequency of the laser
field.

From Eq~(\ref{38}), we find that the dipole matrix elements between the
dressed states and the ground state $|{1}\rangle $ are 
\begin{eqnarray}
\mbox{\boldmath $\mu$}_{a1} &=&\mbox{\boldmath $\mu$}_{13}\sin \phi , 
\nonumber \\
\mbox{\boldmath $\mu$}_{b1} &=&\mbox{\boldmath $\mu$}_{13}\cos \phi .
\label{40}
\end{eqnarray}
Thus, the sub-system with the upper dressed states $|{a}\rangle ,|{b}\rangle 
$ and the ground state $|{1}\rangle $ behaves as a $V$-type system with
parallel dipole moments. This system has the advantage that the magnitudes of
the transition dipole moments and the upper level splitting can be
controlled by the Rabi frequency and detuning of the driving laser field.

\subsection{Two-level atom in a polychromatic field}

\label{sec4b}

Transitions with parallel or anti-parallel dipole moments can be created not
only in multi-level atoms, but also in a two-level atom driven by a
polychromatic field~\cite{ff}. In order to show this, we consider a
two-level atom driven by a bichromatic field composed of a strong resonant
laser field and a weaker laser field detuned from the atomic resonance by
the Rabi frequency of the strong field. The effect of the strong field alone
is to produce dressed states \cite{cohen} 
\begin{eqnarray}
\left|1, N\right\rangle &=& \frac{1}{\sqrt{2}}\left(\left|g, N\right\rangle
- \left|e, N-1\right\rangle \right) ,  \nonumber \\
\left|2, N\right\rangle &=& \frac{1}{\sqrt{2}}\left(\left|g, N\right\rangle
+\left|e, N-1\right\rangle \right) ,  \label{41}
\end{eqnarray}
with energies $E_{1,2}= \hbar\left(N\omega_{0} \pm \frac{1}{2}\Omega\right)$%
, where $N$ is the number of photons in the field mode, $\Omega$ is the Rabi
frequency, and $\omega_{0}$ is the atomic transition frequency.

The dressed states are shown in Fig.~2(a). We see that in the dressed atom
basis the system is no longer a two-level system. It is a multi-level system
with three different transition frequencies, $\omega _{0}$ and $\omega
_{0}\pm \Omega $, and four nonvanishing dipole moments $\mbox{\boldmath
$\mu$}_{ij,N}=\left\langle N,i\right| \mbox{\boldmath $\mu$}\left|
j,N-1\right\rangle $ connecting dressed states between neighbouring
manifolds 
\begin{equation}
\mbox{\boldmath $\mu$}_{11,N}=\mbox{\boldmath $\mu$}_{12,N}=-\mbox{\boldmath
$\mu$}_{21,N}=-\mbox{\boldmath $\mu$}_{22,N}=\frac{1}{2}\mbox{\boldmath
$\mu$}.  \label{42}
\end{equation}

\begin{figure}
\begin{center}
\includegraphics[height=3in,width=5in]{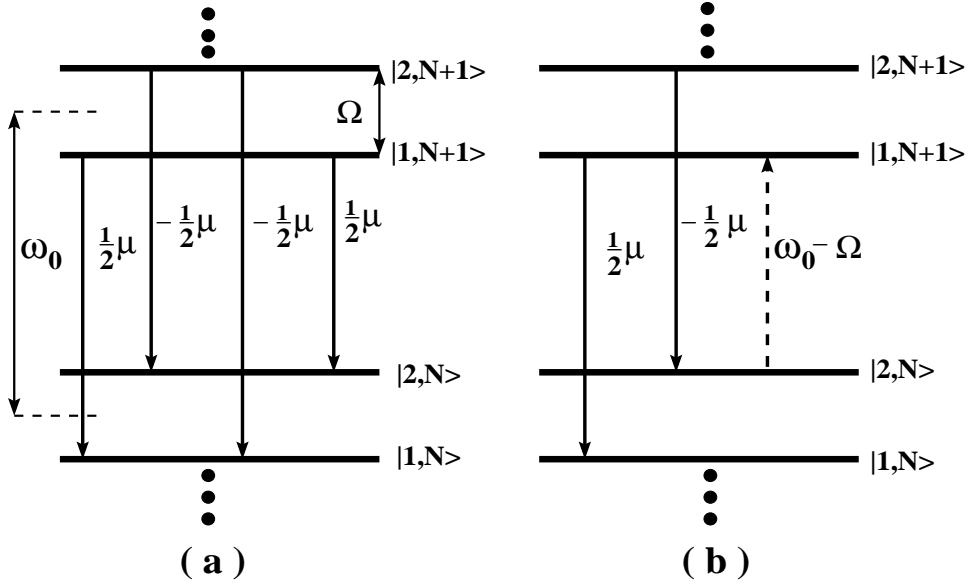}
\caption{(a) Dressed states of a strongly driven two-level atom.
The arrows indicate the allowed spontaneous transitions with dipole
moments $\pm \frac{1}{2}\mu$. (b) A second coherent field (dashed
arrow) of frequency $\omega_{0}-\Omega$ couples the dipole moments of the two degenerate transitions at $\omega_{0}$.}

\end{center}
\end{figure}
There are two transitions with antiparallel dipole moments, $%
\mbox{\boldmath
$\mu$}_{11,N}$ and $\mbox{\boldmath $\mu$}_{22,N}$, that oscillate with the
same frequency $\omega _{0}$. This makes the system an ideal candidate for
quantum interference. However, they are not coupled (correlated), preventing
these dipole moments from being a source of quantum interference. This can
be shown by calculating the correlation functions of the dipole moment
operators of the dressed-atom transitions $\sigma _{ijN}^{+}=\left|
i,N\right\rangle \left\langle N-1,j\right| ,\left( i,j=1,2\right) $. The
correlation functions $\left\langle \sigma _{iiN}^{+}\sigma
_{jjN}^{-}\right\rangle $, $(i\neq j)$, are equal to zero, showing that the
dipole moments oscillate independently.

In order to correlate them, we introduce a second (weaker) laser field of
frequency $\omega _{0}-\Omega $ and Rabi frequency $\Omega _{2}<\Omega $,
which couples the degenerate transitions with dipole moments $%
\mbox{\boldmath $\mu$}_{11,N}$ and $\mbox{\boldmath $\mu$}_{22,N-1}$, as
indicated in Fig.~2(b). Treating the second field perturbatively, at zeroth
order the coupling results in new ``doubly-dressed'' states~\cite{ff} 
\begin{equation}
\left| \bar{N},n\pm \right\rangle =\frac{1}{\sqrt{2}}\left( \left|
2,N-n-1,M+n+1\right\rangle \pm \left| 1,N-n,M+n\right\rangle \right) ,
\label{43}
\end{equation}
where $M$ is the number of photons in the weaker field mode, and $\bar{N}%
=N+M $ is the total number of photons.

On calculating the transition dipole moments $\mbox{\boldmath $\mu$}_{n\pm
,n\pm }$ between the doubly-dressed states, corresponding to the transitions
at $\omega _{0}$, we find that the dipole moments are equal to zero. Thus,
in the doubly-driven atom the effective dipole moments at $\omega _{0}$ are
zero due to quantum interference between the two degenerate dipole moments
of opposite phases. A consequence of this cancellation is the disappearance
of the central component in the fluorescence spectrum of the doubly driven
two-level atom~\cite{ff}.

\subsection{Pre-selected polarization method}

\label{sec4c}

Patnaik and Agarwal~\cite{pa} have proposed a method of generating a
non-zero cross-damping rate in a three-level atom with perpendicular dipole
moments which interacts with a single-mode cavity of a pre-selected
polarization. In this system the polarization index $s$ of the cavity mode
is fixed to only one of the two possible directions. This arrangement of the
polarization can lead to a non-zero cross-damping term $\Gamma _{12}$ in the
master equation of the system, even if the dipole moments of the atomic
transitions are perpendicular. If the polarization of the cavity field is
fixed, say ${\bf e}_{ks}={\bf e}_{kx}$, the polarization direction along the 
$x$-quantization axis, then the cross-damping rate~(\ref{27}) is given by 
\begin{equation}
\Gamma _{12}=\sqrt{\Gamma _{1}\Gamma _{2}}\cos \theta _{1}\cos \theta _{2},
\label{44}
\end{equation}
where $\theta _{j}$ is the angle between $\mbox{\boldmath $\mu$}_{j}$ and
the preselected polarization vector, and usually $\theta _{1}+\theta
_{2}=\pi /2$.

Zhou and Swain~\cite{zs} have shown that the idea of the pre-selected
polarization can be applied to engineer a system with parallel or
anti-parallel dipole moments. Zhou~\cite{zhou} has extended the method to a
cascade three-level atom coupled to a frequency-tunable cavity mode in a
thermal state.

\subsection{Anisotropic vacuum approach}

\label{sec4d}

Agarwal~\cite{ag00} has proposed a totally different mechanism to produce
correlations between two perpendicular dipole moments. In this method the
interference between perpendicular dipole moments is induced by an
anisotropic vacuum field. Using second order perturbation theory, it is
shown that transition probability from the ground state $|{g}\rangle $ of a
four-level system to the final state $|{f}\rangle $ through two intermediate
states $|{i}\rangle $ and $|{j}\rangle $ is given by 
\begin{equation}
T_{gf}=\frac{1}{\hbar ^{2}}\sum_{i,j}\Omega _{i}\Omega _{j}\frac{%
\mbox{\boldmath $\mu$}_{fj}^{\ast }{\bf C}\left( \omega _{L}-\omega
_{fg}\right) \mbox{\boldmath $\mu$}_{fi}}{\left( \omega _{ig}-\omega
_{L}\right) \left( \omega _{jg}-\omega _{L}\right) },  \label{45}
\end{equation}
where $\Omega _{i}(\Omega _{j})$ is the Rabi frequency of the $|{g}\rangle
\rightarrow |{i}\rangle (|{g}\rangle \rightarrow |{j}\rangle )$ transition, $%
\omega _{L}$ is the frequency of the driving laser, and ${\bf C}\left(
\omega _{L}-\omega _{fg}\right) $ is the Fourier transform of the tensor,
anti-normally ordered correlation function of the vacuum field operators.
The anisotropy of the vacuum enters through the tensor ${\bf C}$. With
perpendicular dipole moments $\mbox{\boldmath $\mu$}_{fj}$ and $%
\mbox{\boldmath $\mu$}_{fi}$, the transition probability responsible for the
quantum interference between the $|{i}\rangle \rightarrow |{f}\rangle $ and $%
|{j}\rangle \rightarrow |{f}\rangle $ transitions may be non-zero only if
the tensor ${\bf C}$ is anisotropic. For an isotropic vacuum the tensor $%
{\bf C}$ is proportional to the unit tensor and then the transition
probability vanishes for $\mbox{\boldmath $\mu$}_{fi}\perp 
\mbox{\boldmath
$\mu$}_{fj}$.

\section{Quantum interference in a two-atom system}

\label{sec5}

In the Young's interference experiment the slits can be replaced by two
atoms and interference effects can be observed between the coherent or
incoherent fields emitted from the atoms. The advantage of using atoms
instead of slits is that at a given time each atom cannot emit more than one
photon. Therefore, the atoms can be regarded as sources of single photon
fields.

Many interesting interference effects have been predicted in the
fluorescence field emitted from two atoms, and the interference fringes have
been observed experimentally in the resonance fluorescence of two trapped
ions~\cite{eich}. In the theoretical analysis various systems have been
considered including non-identical atoms~\cite{ftk98}, the effect of
interatomic interactions~\cite{du}, and the dependence of the interference
pattern on the direction of propagation of a driving field with respect to
the interatomic axis~\cite{fs}.

Here, we derive general criteria for the first- and second-order
interference in the fluorescence field emitted from two two-level atoms.
Using these criteria, we may easily predict conditions for quantum
interference in the two atom system. We consider two atoms with the upper
level $|e_{i}\rangle $ and the ground level $|g_{i}\rangle $ $(i=1,2),$
located at points ${\bf r}_{1}$ and ${\bf r}_{2}$. The atoms can have
identical or non-identical frequencies $\omega _{i}$ and can be coupled
through the vacuum field. Moreover, the atoms may be driven by arbitrary,
external fields.

Using Eq.~(\ref{24}), which relates the electric field operator to the
atomic dipole operators, we obtain the following expressions for the time
and the angular distribution of the first- and second-order correlations 
\begin{eqnarray}
G^{\left( 1\right) }\left( {\bf R},t\right) &=& u\left({\bf R}\right)
\sum_{i,j=1}^{2}\left(\Gamma_{i}\Gamma_{j}\right)^{\frac{1}{2}}\left\langle
S_{i}^{+}\left( t\right) S^{-}_{j}\left( t\right) \right\rangle \exp \left(ik%
\bar{{\bf R}}\cdot {\bf r}_{ij}\right) ,  \label{46} \\
G^{\left( 2\right)}\left( {\bf R},t;{\bf R},t\right) &=& u\left({\bf R}%
_{1}\right)u\left({\bf R}_{2}\right) \sum_{i,j,k,l=1}^{2}\left(\Gamma
_{i}\Gamma_{j}\Gamma _{k}\Gamma_{l}\right)^{\frac{1}{2}}  \nonumber \\
&\times& \left\langle S^{+}_{i}\left( t\right) S^{+}_{k}\left(t \right)
S^{-}_{l}\left( t\right) S^{-}_{j}\left(t\right) \right\rangle \exp \left[%
ik\left(\bar{{\bf R}}_{1}\cdot {\bf r}_{ij}+\bar{{\bf R}}_{2} \cdot {\bf r}%
_{kl}\right)\right] ,  \label{47}
\end{eqnarray}
where $u\left({\bf R}\right)$ is a constant which depends on the geometry of
the system.

The traditional method to analyse coherence properties of light emitted from
two atoms is to examine specific processes, such as spontaneous emission or
resonance fluorescence of driven atoms. In this approach equations of motion
are derived for the atomic correlation functions appearing in Eqs.~(\ref{46}%
) and (\ref{47}), and solved using standard mathematical methods. Here, we
present an alternative approach which allows us to identify general
conditions for the observation of coherence effects without examining
specific processes. In this approach, we introduce the collective states
(Dicke states) of the two-atom system~\cite{dicke} 
\begin{eqnarray}
\left| g\right\rangle &=&\left| g_{1}\right\rangle \left| g_{2}\right\rangle
,  \nonumber \\
\left| s\right\rangle &=&\frac{1}{\sqrt{2}}\left( \left| g_{1}\right\rangle
\left| e_{2}\right\rangle +\left| e_{1}\right\rangle \left|
g_{2}\right\rangle \right) ,  \nonumber \\
\left| a\right\rangle &=&\frac{1}{\sqrt{2}}\left( \left| g_{1}\right\rangle
\left| e_{2}\right\rangle -\left| e_{1}\right\rangle \left|
g_{2}\right\rangle \right) ,  \nonumber \\
\left| e\right\rangle &=&\left| e_{1}\right\rangle \left| e_{2}\right\rangle
,  \label{48}
\end{eqnarray}
where $|s\rangle $ and $|a\rangle $ are the entangled symmetric and
antisymmetric atomic collective states, respectively.

In the basis of the collective states~(\ref{48}) the atomic correlation
functions, appearing in Eqs.~(\ref{46}) and (\ref{47}), are given by 
\begin{eqnarray}
\left\langle S_{1}^{+}S_{1}^{-}\right\rangle + \left\langle
S_{2}^{+}S_{2}^{-}\right\rangle &=& \rho_{ss}+\rho_{aa}+2\rho_{ee} , 
\nonumber \\
\left\langle S_{1}^{+}S_{2}^{-}\right\rangle &=& \frac{1}{2}%
\left(\rho_{ss}-\rho_{aa}+\rho_{as}-\rho_{sa}\right) ,  \nonumber \\
\left\langle S_{1}^{+}S_{2}^{+}S_{1}^{-}S_{2}^{-}\right\rangle &=& \rho_{ee}
,  \label{49}
\end{eqnarray}
where $\rho_{ii} (i=a,s,e)$ are the populations of the collective states and 
$\rho_{sa}, \rho_{as}$ are coherences.

Using the relations~(\ref{49}), we find 
\begin{eqnarray}
G^{\left( 1\right) }\left( {\bf R},t\right)  &=&\Gamma u\left( {\bf R}%
\right) \left\{ 2\rho _{ee}\left( t\right) +\rho _{ss}\left( t\right) \left(
1+\cos k\bar{{\bf R}}\cdot {\bf r}_{21}\right) \right.   \nonumber \\
&+&\left. \rho _{aa}\left( t\right) \left( 1-\cos k\bar{{\bf R}}\cdot {\bf r}%
_{21}\right) \right.   \nonumber \\
&+&\left. i\left( \rho _{sa}\left( t\right) -\rho _{as}\left( t\right)
\right) \sin k\bar{{\bf R}}\cdot {\bf r}_{21}\right\} ,  \label{50}
\end{eqnarray}
and 
\begin{equation}
G^{\left( 2\right) }\left( {\bf R},t;{\bf R},t\right) =4\Gamma ^{2}u\left( 
{\bf R}_{1}\right) u\left( {\bf R}_{2}\right) \rho _{ee}\left( t\right) %
\left[ 1+\cos k\left( \bar{{\bf R}}_{1}-\bar{{\bf R}}_{2}\right) \cdot {\bf r%
}_{21}\right] .  \label{51}
\end{equation}
It is evident from Eq.~(\ref{50}) that the first-order correlation function
can exhibit an interference pattern only if $\rho _{ss}\neq \rho _{aa}$
and/or Im$(\rho _{sa})\neq 0$. This happens when $\langle e_{1}|\langle
g_{2}|\rho |e_{2}\rangle |g_{1}\rangle $ and $\langle g_{1}|\langle
e_{2}|\rho |g_{2}\rangle |e_{1}\rangle $ are different from zero, i.e. when
there are non-zero coherences between the atoms. On the other hand, the
second-order correlation function is independent of the populations $\rho
_{ss},\rho _{aa}$ and the coherences, and exhibits an interference pattern
when $\rho _{ee}(t)\neq 0$.

We now examine some specific processes in which one can create unequal
populations of the $|s\rangle $ and $|a\rangle $ states. Dung and Ujihara~ 
\cite{du} have shown that spontaneous emission from two identical atoms,
with initially only one atom excited, can exhibit an interference pattern.
It is easy to interpret this effect in terms of the populations $\rho
_{ss}(t)$ and $\rho _{aa}(t)$. If initially only one atom is excited; $\rho
_{ee}(0)=\rho _{sa}(0)=\rho _{as}(0)=0$ and $\rho _{ss}(0)=\rho _{aa}(0)=%
\frac{1}{2}$. Using the master equation~(\ref{29}), we find that the time
evolution of the populations $\rho _{ss}(t)$ and $\rho _{aa}(t)$ is given by 
\begin{eqnarray}
\rho _{ss}\left( t\right) &=&\rho _{ss}\left( 0\right) \exp \left[ -\left(
\Gamma +\Gamma _{12}\right) t\right] ,  \nonumber \\
\rho _{aa}\left( t\right) &=&\rho _{aa}\left( 0\right) \exp \left[ -\left(
\Gamma -\Gamma _{12}\right) t\right] .  \label{52}
\end{eqnarray}
Since the populations decay with different rates, the symmetric state decays
with an enhanced rate $\Gamma +\Gamma _{12}$, while the antisymmetric state
decays with a reduced rate $\Gamma -\Gamma _{12}$, and the population $\rho
_{aa}(t)$ is larger than $\rho _{ss}(t)$ for all $t>0$. Hence, an
interference pattern can be observed for $t>0$. This effect arises from the
presence of the interatomic interactions $(\Gamma _{12}\neq 0)$. Thus, for
two independent atoms the populations decay with the same rate resulting in
the disappearance of the interference pattern.

When the atoms are driven by a coherent laser field, an interference pattern
can be observed even in the absence of the interatomic interactions. To show
this, we find from the master equation~(\ref{29}) the steady-state solutions
for the populations of the collective atomic states 
\begin{eqnarray}
\rho _{ee} &=&\frac{\Omega ^{4}}{4D},  \nonumber \\
\rho _{ss} &=&\frac{2\Omega ^{2}\left( \Gamma ^{2}+\Delta _{L}^{2}\right)
+\Omega ^{4}}{4D},  \nonumber \\
\rho _{aa} &=&\frac{\Omega ^{4}}{4D},  \label{53}
\end{eqnarray}
where $\Omega $ is the Rabi frequency of the driving field, 
\begin{equation}
D=\Omega ^{4}+\left( \Gamma ^{2}+\Delta _{L}^{2}\right) \left\{ \Omega ^{2}+%
\frac{1}{4}\left[ \left( \Gamma +\Gamma _{12}\right) ^{2}+\left( \Delta
_{L}-\Omega _{12}\right) ^{2}\right] \right\} ,  \label{54}
\end{equation}
and $\Delta _{L}=\omega _{0}-\omega _{L}$ is the detuning of the laser
frequency from the atomic transition frequency. In the derivation of Eq.~(%
\ref{53}), we have assumed that the laser field propagates in the direction
perpendicular to the interatomic axis $({\bf k}\cdot {\bf r}_{21})$ such
that both atoms experience the same driving field amplitude and phase.

It is evident from Eq.~(\ref{53}) that $\rho_{ss}>\rho_{aa}$ even in the
absence of the interatomic interactions $(\Gamma_{12}=\Omega_{12}=0)$.
Hence, an interference pattern can be observed even for two independent
atoms. In this case the interference pattern results from the coherent
synchronization of the oscillations of the atoms by the constant coherent
phase of the driving laser field.

We have shown that first-order coherence is sensitive to the interatomic
interactions and the excitation field. In contrast, the second-order
correlation function can exhibit an interference pattern independent of the
interatomic interactions and the excitation process. According to Eq.(\ref
{51}), to observe an interference pattern in the second-order correlation
function, it is enough to produce a non-zero population in the state $%
|e\rangle $. This effect results from the detection process, in that a
detector does not distinguish between two simultaneously detected photons.

\section{Quantum interference as a control of spontaneous emission}

\label{sec6}

The master equation~(\ref{29}) shows that quantum interference modifies
spontaneous emission rates from an atomic system. The modification and
control of spontaneous emission is a topic of much current interest because
of the many possible applications in quantum computation and quantum
information theory. As spontaneous emission arises from the interaction of
an atomic system with the vacuum field, the most obvious mechanism for
modifying spontaneous emission is to place the system in a reservoir such as
an electromagnetic cavity, an optical waveguide, or a photonic band-gap
material. These reservoirs change the density of modes of the vacuum field
into which the system can emit. For atoms in free space, quantum
interference has been recognized as the basic phenomenon for controlling
spontaneous emission. It was first shown by Agarwal~\cite{ag74} that the
decay of an excited degenerate $V$-type three-level atom can be modified due
to interference between the two coupled atomic transitions, and a population
trapping can occur.

\subsection{Vacuum-induced superposition systems}

\label{sec6a}

The modification of spontaneous emission from two-atom or two-channel
systems results from the quantum interference induced linear superpositions
of the atomic transitions. To show this, we introduce superposition
operators which are linear combinations of the bare atomic dipole operators~ 
\cite{afs} 
\begin{eqnarray}
S_{s}^{+} &=&uS_{1}^{+}+vS_{2}^{+},  \nonumber \\
S_{a}^{+} &=&vS_{1}^{+}-uS_{2}^{+},  \label{55}
\end{eqnarray}
where 
\begin{equation}
u=\frac{\sqrt{\Gamma _{1}}}{\sqrt{\Gamma _{1}+\Gamma _{2}}},\qquad v=\frac{%
\sqrt{\Gamma _{2}}}{\sqrt{\Gamma _{1}+\Gamma _{2}}},  \label{56}
\end{equation}
and $\left| u\right| ^{2}+\left| v\right| ^{2}=1$.

The operators $S_{s}^{+}$ and $S_{a}^{+}$ represent, respectively, symmetric
and antisymmetric combinations of the dipole moments of the two systems. In
terms of these new operators, the damping part of the master equation~(\ref
{29}) may be written as 
\begin{eqnarray}
{\cal {L}}_{d}\rho &=&-\Gamma _{ss}\left( S_{s}^{+}S_{s}^{-}\rho +\rho
S_{s}^{+}S_{s}^{-}-2S_{s}^{-}\rho S_{s}^{+}\right)  \nonumber \\
&&-\Gamma _{aa}\left( S_{a}^{+}S_{a}^{-}\rho +\rho
S_{a}^{+}S_{a}^{-}-2S_{a}^{-}\rho S_{a}^{+}\right)  \nonumber \\
&&-\Gamma _{sa}\left( S_{s}^{+}S_{a}^{-}\rho +\rho
S_{s}^{+}S_{a}^{-}-2S_{a}^{-}\rho S_{s}^{+}\right)  \nonumber \\
&&-\Gamma _{as}\left( S_{a}^{+}S_{s}^{-}\rho +\rho
S_{a}^{+}S_{s}^{-}-2S_{s}^{-}\rho S_{a}^{+}\right) ,  \label{57}
\end{eqnarray}
where 
\begin{eqnarray}
\Gamma _{ss} &=&\frac{1}{2}\frac{\left( \Gamma _{1}^{2}+\Gamma
_{2}^{2}+2\Gamma _{12}\sqrt{\Gamma _{1}\Gamma _{2}}\right) }{\Gamma
_{1}+\Gamma _{2}},  \nonumber \\
\Gamma _{aa} &=&\frac{\left( \sqrt{\Gamma _{1}\Gamma _{2}}-\Gamma
_{12}\right) \sqrt{\Gamma _{1}\Gamma _{2}}}{\Gamma _{1}+\Gamma _{2}}, 
\nonumber \\
\Gamma _{sa} &=&\Gamma _{as}=\frac{1}{2}\frac{\left( \Gamma _{1}-\Gamma
_{2}\right) \left( \sqrt{\Gamma _{1}\Gamma _{2}}-\Gamma _{12}\right) }{%
\Gamma _{1}+\Gamma _{2}}.  \label{58}
\end{eqnarray}

Although in general the two forms (\ref{29}) and (\ref{57}) look similar,
the advantage of the transformed form (\ref{57}) over (\ref{29}) is that the
damping rates of the superposition systems are significantly different even
if the damping rates of the original systems are equal. When $\Gamma
_{1}=\Gamma _{2}$ the damping rates satisfy $\Gamma _{sa}=\Gamma _{as}=0$,
and then the symmetric and antisymmetric superpositions decay independently
with rates $\frac{1}{2}\left( \Gamma +\Gamma _{12}\right) $ and $\frac{1}{2}%
\left( \Gamma -\Gamma _{12}\right) $ respectively. In other words, for $%
\Gamma _{1}=\Gamma _{2}$ the transformation~(\ref{55}) diagonalizes the
dispersive part of the master equation. Furthermore, if $\Gamma _{12}=\sqrt{%
\Gamma _{1}\Gamma _{2}}$ then $\Gamma _{aa}=\Gamma _{sa}=\Gamma _{as}=0$
regardless of the ratio between $\Gamma _{1}$ and $\Gamma _{2}$. In this
case the antisymmetric combination does not decay. This implies that
spontaneous emission can be controlled and even suppressed by appropriately
engineering the cross-damping rate $\Gamma _{12}$.

\subsection{Population trapping and dark states}

\label{sec6b}

In the literature, population trapping is often referred to as a consequence
of the cancellation of spontaneous emission. Here, we point out that the
cancellation of spontaneous emission from an atomic state does not always
lead to the trapping of the population in this non-decaying state. We
illustrate this by considering the process of spontaneous emission from a $V$%
-type atom. For simplicity, we assume that spontaneous emission occurs from
the excited states to the ground state with the same decay rates $\Gamma
_{1}=\Gamma _{2}=\Gamma $, and the transition between the excited states is
forbidden in the electric dipole approximation. The allowed transitions are
represented by the dipole operators $S_{1}^{+}=\left( S_{1}^{-}\right)
^{\dagger }=\left| 1\right\rangle \left\langle 2\right| $ and $%
S_{2}^{+}=\left( S_{2}^{-}\right) ^{\dagger }=\left| 3\right\rangle
\left\langle 2\right| $. In the absence of the driving field $(\Omega
_{1}=\Omega _{2}=0)$, the master equation~(\ref{29}) leads to the following
equations of motion for the density matrix elements 
\begin{eqnarray}
\dot{\rho}_{11} &=&-\Gamma \rho _{11}-\frac{1}{2}\Gamma _{12}\left( \rho
_{13}+\rho _{31}\right) ,  \nonumber \\
\dot{\rho}_{33} &=&-\Gamma \rho _{33}-\frac{1}{2}\Gamma _{12}\left( \rho
_{13}+\rho _{31}\right) ,  \nonumber \\
\dot{\rho}_{22} &=&\Gamma \left( \rho _{11}+\rho _{33}\right) +\Gamma
_{12}\left( \rho _{13}+\rho _{31}\right) ,  \nonumber \\
\dot{\rho}_{13} &=&-\left( \Gamma +i\Delta \right) \rho _{13}-\frac{1}{2}%
\Gamma _{12}\left( \rho _{11}+\rho _{33}\right) ,  \nonumber \\
\dot{\rho}_{31} &=&-\left( \Gamma -i\Delta \right) \rho _{31}-\frac{1}{2}%
\Gamma _{12}\left( \rho _{11}+\rho _{33}\right) ,  \label{59}
\end{eqnarray}
where $\Delta =\omega _{1}-\omega _{2}$ is the detuning between the atomic
transitions and, for simplicity, we have ignored the small frequency shifts $%
\delta _{12}^{(\pm )}$.

There are two different steady-state solutions of Eq.~(\ref{59}) depending
on whether the transitions are degenerate $(\Delta=0)$ or non-degenerate $%
(\Delta \neq 0)$. This fact is connected with the existence of a linear
combination of the density matrix elements 
\begin{equation}
\alpha\left(t\right) = \rho_{11}\left(t\right)+\rho_{33}\left(t\right)
-\rho_{13}\left(t\right)-\rho_{31}\left(t\right) ,  \label{60}
\end{equation}
which, for $\Delta=0$ and $\Gamma_{12}=\Gamma$ is a constant of motion~\cite
{ag74,buz}.

In the case of $\Delta=0$ and $\Gamma_{12}=\Gamma$, the steady-state
solution of Eq.~(\ref{59}) is 
\begin{eqnarray}
\rho_{11}\left(\infty\right) &=& \rho_{33}\left(\infty\right) =\frac{1}{4}%
\alpha\left(0\right) ,  \nonumber \\
\rho_{13}\left(\infty\right) &=& \rho_{31}\left(\infty\right) =-\frac{1}{4}%
\alpha\left(0\right) ,  \nonumber \\
\rho_{22}\left(\infty\right) &=&\frac{1}{2}\alpha\left(0\right) .  \label{61}
\end{eqnarray}
It is seen that the steady-state population distribution depends on the
initial population. When $\alpha\left(0\right)\neq 0$ a part of the
population can remain in the excited states.

On the other hand, for $\Delta \neq 0$ and/or $\Gamma _{12}\neq \Gamma ,$
the linear combination (\ref{60}) is no longer a constant of the motion, and
then the steady-state solution of Eq.~(\ref{59}) is 
\begin{eqnarray}
\rho _{11}\left( \infty \right) &=&\rho _{33}\left( \infty \right) =\rho
_{13}\left( \infty \right) =\rho _{31}\left( \infty \right) =0,  \nonumber \\
\rho _{22}\left( \infty \right) &=&1.  \label{62}
\end{eqnarray}
In this case the population distribution does not depend on the initial
state of the atom and in the steady-state the population is in the ground
state.

The properties of this system can be understood by transforming to new
states which are linear superpositions of the excited atomic states 
\begin{eqnarray}
|{s}\rangle &=&\frac{1}{\sqrt{2}}\left( |{1}\rangle +|{3}\rangle \right) , 
\nonumber \\
|{a}\rangle &=&\frac{1}{\sqrt{2}}\left( |{1}\rangle -|{3}\rangle \right) .
\label{63}
\end{eqnarray}
From Eqs.~(\ref{59})and (\ref{63}), we find the following equations of
motion for the populations of these superposition states 
\begin{eqnarray}
\dot{\rho}_{ss} &=&-\frac{1}{2}\left( \Gamma +\Gamma _{12}\right) \rho _{ss}-%
\frac{1}{2}i\Delta \left( \rho _{sa}-\rho _{as}\right) ,  \nonumber \\
\dot{\rho}_{aa} &=&-\frac{1}{2}\left( \Gamma -\Gamma _{12}\right) \rho _{aa}+%
\frac{1}{2}i\Delta \left( \rho _{sa}-\rho _{as}\right) .  \label{64}
\end{eqnarray}
It is seen that the antisymmetric state decays at a reduced rate $\left(
\Gamma -\Gamma _{12}\right) $, and for $\Gamma _{12}=\Gamma $ the state does
not decay at all. In this case the antisymmetric state can be regarded as a 
{\it dark state} in the sense that the state is decoupled from the
environment. Secondly, we note from Eq.~(\ref{64}) that the population
oscillates between the states with the amplitude $\Delta $, which plays here
a role similar to that of the Rabi frequency of the coherent interaction
between the symmetric and antisymmetric states. Consequently, an initial
population in the state $|{a}\rangle $ can be coherently transferred to the
state $|{s}\rangle $, which decays rapidly to the ground state. When $\Delta
=0$, the coherent interaction does not take place and then any initial
population in $|{a}\rangle $ will stay in this state for all times. In this
case we can say that the population is {\it trapped} in the state $|{a}%
\rangle $.

We conclude that cancellation of spontaneous emission does not necessarily
lead to population trapping. The population can be trapped in a dark state
only if the state is completely decoupled from any interactions.

\section{Quantum interference effects in coherently driven systems}

\label{sec7}

In the preceding section, we discussed the effect of quantum interference on
spontaneous emission in a two-channel system. By means of specific examples
we have demonstrated that spontaneous emission can be controlled and even
suppressed by quantum interference. In this section, we extend the analysis
to the case of coherently driven systems. We will focus on the effect of
quantum interference on transition rates between dressed states of the
system. In particular, we consider coherently driven $V$ and $\Lambda $-type
three-level atoms.

\subsection{Excitation from an auxiliary level}

\label{sec7a}

Our first example for quantum interference in driven atomic systems is a
three-level atom in the $V$ configuration composed of two non-degenerate
excited states $\left| 1\right\rangle $ and $\left| 3\right\rangle $ and a
single ground state $\left| 2\right\rangle $. As before, we assume that the
upper states $\left| 1\right\rangle $ and $\left| 3\right\rangle $ decay to
the ground state by spontaneous emission with decay rates $\Gamma _{1}$ and $%
\Gamma _{2}$, respectively, whereas transitions between the excited levels
are forbidden in the electric dipole approximation. The allowed transitions
have dipole moments $\mbox{\boldmath $\mu$}_{12}$ and $\mbox{\boldmath $\mu$}%
_{32}$ sharing the same ground state $\left| 2\right\rangle ,$ and are
represented by the operators $S_{1}^{+}=\left( S_{1}^{-}\right) ^{\dagger
}=\left| 1\right\rangle \left\langle 2\right| $ and $S_{2}^{+}=\left(
S_{2}^{-}\right) ^{\dagger }=\left| 3\right\rangle \left\langle 2\right| $.
The transitions may be driven by a coherent laser field from an auxiliary
level or the laser field may couple directly to the decaying transitions.

Zhu and Scully~\cite{zsc} have shown that quantum interference in a $V$-type
system, driven by a laser field from an auxiliary level, can lead to the
elimination of the spectral line at the driving laser frequency. The
four-level system considered by Zhu and Scully is shown in Fig.~3. The laser
field is coupled to non-decaying $|{1}\rangle -|{b}\rangle $ and $|{3}%
\rangle -|{b}\rangle $ transitions, whereas spontaneous emission occurs from
the levels $|{1}\rangle $ and $|{3}\rangle $ to the ground level $|{2}%
\rangle $.

\begin{figure}
\begin{center}
\includegraphics[height=2in,width=4.5in]{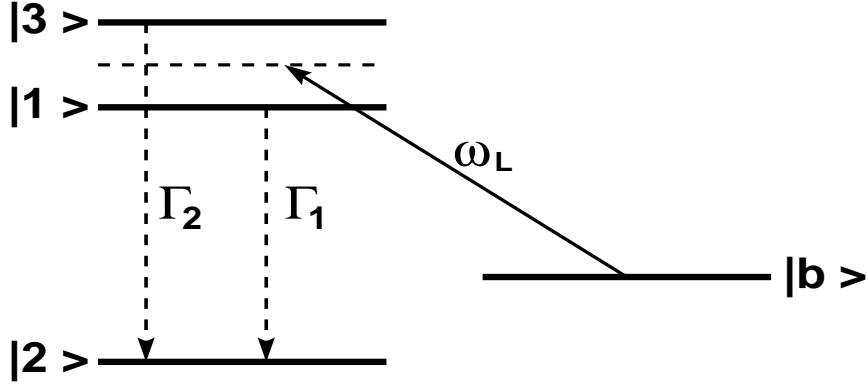}
\caption{Three-level $V$-type system driven from an auxiliary level.}
\end{center}
\end{figure}

The most direct approach to the analysis of the dynamics of the system is
the master equation~(\ref{29}) with the Hamiltonian $H^{\prime }$ given by 
\begin{eqnarray}
H^{\prime } &=&\hbar \omega _{1}S_{1}^{+}S_{1}^{-}+\hbar \omega
_{2}S_{2}^{+}S_{2}^{-}+\hbar \omega _{b}|b\rangle \langle b|  \nonumber \\
&-&\frac{1}{2}\hbar \left[ \left( \Omega _{1}S_{1b}^{+}+\Omega
_{2}S_{3b}^{+}\right) \exp \left( -i\omega _{L}t\right) +{\rm H.c.}\right] ,
\label{65}
\end{eqnarray}
where $S_{1b}^{+}=|{1}\rangle \langle {b}|$ and $S_{3b}^{+}=|{3}\rangle
\langle {b}|$ are the dipole raising operators for the transitions between
the upper levels $|{1}\rangle $ and $|{3}\rangle $ and the auxiliary level $|%
{b}\rangle $.

The spectrum of the fluorescence field emitted on the $|{1}\rangle
\rightarrow |{2}\rangle $ and $|{3}\rangle \rightarrow |{2}\rangle $
transitions is given by the Fourier transform of the two-time correlation
function of the dipole moments of the transitions that, according to the
quantum regression theorem~\cite{lax}, satisfy the same equations of motion
as the density matrix elements $\rho _{12}(t)$ and $\rho _{32}(t)$. Using
the master equation~(\ref{28}) with the Hamiltonian~(\ref{65}), we obtain
the following set of coupled equations of motion for the density matrix
elements 
\begin{equation}
\frac{\partial }{\partial t}{\bf X}\left( t\right) =M{\bf X}\left( t\right) ,
\label{66}
\end{equation}
where ${\bf X}\left( t\right) =[\rho _{12}(t),\rho _{32}(t),\rho _{b2}(t)]$
is a column vector composed of the density matrix elements, and $M$ is the $%
3\times 3$ matrix 
\begin{equation}
M=\left( 
\begin{array}{ccc}
-(\frac{1}{2}\Gamma _{1}+i\Delta _{1}) & -\frac{1}{2}\Gamma _{12} & \frac{1}{%
2}i\Omega _{1} \\ 
-\frac{1}{2}\Gamma _{12} & -(\frac{1}{2}\Gamma _{2}+i\Delta _{2}) & \frac{1}{%
2}i\Omega _{2} \\ 
\frac{1}{2}i\Omega _{1} & \frac{1}{2}i\Omega _{2} & 0
\end{array}
\right) ,  \label{67a}
\end{equation}
where $\Delta _{1}=\omega _{1b}-\omega _{L}$ and $\Delta _{2}=\omega
_{3b}-\omega _{L}$ are the detunings of the laser field from the $|{1}%
\rangle -|{b}\rangle $ and $|{3}\rangle -|{b}\rangle $ transitions,
respectively. Following Zhu and Scully, we assume that $\Gamma _{1}=\Gamma
_{2}=\Gamma $, $\Omega _{1}=\Omega _{2}=\Omega $ and that the laser field is
tuned to the middle of the upper levels spitting, i.e. $\Delta _{2}=-\Delta
_{1}=\frac{1}{2}\Delta $.

Since we are interested in the time evolution of the density matrix
elements, we need explicit expressions for the components $X_{i}$ of the
vector ${\bf X}\left( t\right) $ in terms of their initial values. This can
be done by a direct integration of (\ref{66}). Thus, if $t_{0}$ denotes an
arbitrary initial time, the integration of (\ref{66}) leads to the following
formal solution for ${\bf X}\left( t\right) $ 
\begin{equation}
{\bf X}\left( t\right) ={\bf X}\left( t_{0}\right) \exp \left( Mt\right) .
\label{68}
\end{equation}
Because the determinant of the matrix $M$ is different from zero, there
exists a complex invertible matrix $T$ which diagonalises $M$, and $\lambda
=T^{-1}MT$ is the diagonal matrix of complex eigenvalues, which can be found
from the eigenvalue equation 
\begin{equation}
\lambda \left[ \lambda ^{2}+\Gamma \lambda +\frac{1}{4}\Delta ^{2}+\frac{1}{4%
}\left( \Gamma ^{2}-\Gamma _{12}^{2}\right) \right] +\frac{1}{2}\Omega ^{2}%
\left[ \lambda +\frac{1}{2}\left( \Gamma -\Gamma _{12}\right) \right] =0.
\label{69}
\end{equation}
There are two different solutions of Eq.~(\ref{69}) depending on whether $%
\Gamma _{12}=\Gamma $ or $\Gamma _{12}\neq \Gamma $. For $\Gamma
_{12}=\Gamma $, which corresponds to parallel dipole moments of the
transitions, and $\Omega \gg \Gamma $ the roots of the cubic equation~(\ref
{69}) are 
\begin{eqnarray}
\lambda _{1} &=&0,  \nonumber \\
\lambda _{2} &=&-\frac{1}{2}\Gamma +i\Omega ^{\prime },  \nonumber \\
\lambda _{3} &=&-\frac{1}{2}\Gamma -i\Omega ^{\prime },  \label{70}
\end{eqnarray}
whilst for $\Gamma _{12}=0$, which corresponds to perpendicular dipole
moments, and $\Omega \gg \Gamma $ the roots are 
\begin{eqnarray}
\lambda _{1} &=&-\frac{1}{2}\Gamma ,  \nonumber \\
\lambda _{2} &=&-\frac{1}{4}\Gamma +i\Omega ^{\prime },  \nonumber \\
\lambda _{3} &=&-\frac{1}{4}\Gamma -i\Omega ^{\prime },  \label{71}
\end{eqnarray}
where $\Omega ^{\prime }=\frac{1}{2}\sqrt{\Delta ^{2}+2\Omega ^{2}}$.

Thus, in the case of parallel dipole moments, the spectrum is composed of
two lines of equal bandwidths $(\frac{1}{2}\Gamma )$ located at frequencies $%
\pm \Omega ^{\prime }$ and there is no central component in the fluorescence
spectrum at the laser frequency $\omega _{L}$. The eigenvalue $\lambda =0$
contributes to the coherent scattering of the laser field. When $\Gamma
_{12}=0$, the spectrum is composed of three lines: the central line of the
bandwidth $\frac{1}{2}\Gamma $ located at the laser frequency and two
sidebands of bandwidths $\frac{1}{4}\Gamma $ located at $\pm \Omega ^{\prime
}$. The absence of the central line for $\Gamma _{12}=\Gamma $ is clear
evidence of the cancellation of spontaneous emission into the vacuum modes
around the laser frequency by quantum interference.

The physical origin of the cancellation of the central line in the spectrum
can be explained clearly by the dressed-atom model of the system~\cite
{cohen,lee}. In this model we use a fully quantum-mechanical description of
the Hamiltonian of the system, which in a frame rotating with the laser
frequency $\omega_{L}$ can be written as 
\begin{equation}
H^{\prime }=H_{0b}+V_{b},  \label{72}
\end{equation}
where 
\begin{equation}
H_{0b}= \hbar \Delta_{1} S_{1}^{+}S_{1}^{-} +\hbar \Delta_{2}
S_{2}^{+}S_{2}^{-}+\hbar \omega_{L}a_{L}^{\dagger}a_{L} ,  \label{73}
\end{equation}
is the Hamiltonian of the uncoupled system and the laser field, and 
\begin{equation}
V_{b}=-\frac{\hbar }{2}g\left[\left(S_{1b}^{+} +S_{2b}^{+}\right)a_{L} +
a_{L}^{\dagger}\left(S_{1b}^{-}+ S_{2b}^{-}\right)\right]  \label{74}
\end{equation}
is the interaction of the laser with the atom. In Eq.~(\ref{74}), $g$ is the
system-field coupling constant, and $a_{L}$ $(a_{L}^{\dagger })$ is the
annihilation (creation) operator for the driving field mode.

For $\Delta_{2}=-\Delta_{1}=\frac{1}{2}\Delta$, the Hamiltonian $H_{0b}$ has
four non-degenerate eigenstates $\left| 2,N\right\rangle,$ $\left|
b,N\right\rangle $, $\left| 1,N-1\right\rangle $, and $\left|
3,N-1\right\rangle$, where $\left| i,N\right\rangle $ is the state with the
atom in state $\left| i\right\rangle $ and $N$ photons present in the
driving laser mode. When we include the interaction $V_{b}$, the
diagonalization of the Hamiltonian $H_{0b}+V_{b}$ leads to the following
dressed states of the system 
\begin{eqnarray}
\left| +,N\right\rangle &=&\frac{1}{2}\left[\left(1-\alpha\right) \left|
1,N-1\right\rangle + \left(1+\alpha\right) \left| 3,N-1\right\rangle -
2\beta \left| b,N\right\rangle \right] ,  \nonumber \\
\left| 0,N\right\rangle &=&-\beta \left(\left|1,N-1\right\rangle -
\left|3,N-1\right\rangle\right) +\alpha \left| b,N\right\rangle ,  \nonumber
\\
\left| -,N\right\rangle &=&\frac{-1}{2}\left[\left(1+\alpha\right) \left|
1,N-1\right\rangle + \left(1-\alpha\right) \left| 3,N-1\right\rangle +
2\beta \left| b,N\right\rangle \right] ,  \nonumber \\
\left| 2,N\right\rangle &=&\left| 2,N\right\rangle  \label{75}
\end{eqnarray}
with energies 
\begin{eqnarray}
E_{N,+} &=& \hbar \left(N\omega_{L}+\Omega^{\prime}\right) ,  \nonumber \\
E_{N,0} &=& \hbar N\omega_{L} ,  \nonumber \\
E_{N,-} &=& \hbar \left(N\omega_{L}-\Omega^{\prime}\right) ,  \nonumber \\
E_{N,2} &=& \hbar N\omega_{L} ,  \label{76}
\end{eqnarray}
where $\alpha =\Delta /2\Omega ^{\prime }$ and $\beta =\Omega /2\Omega
^{\prime }$.

\begin{figure}
\begin{center}
\includegraphics[height=3in,width=5in]{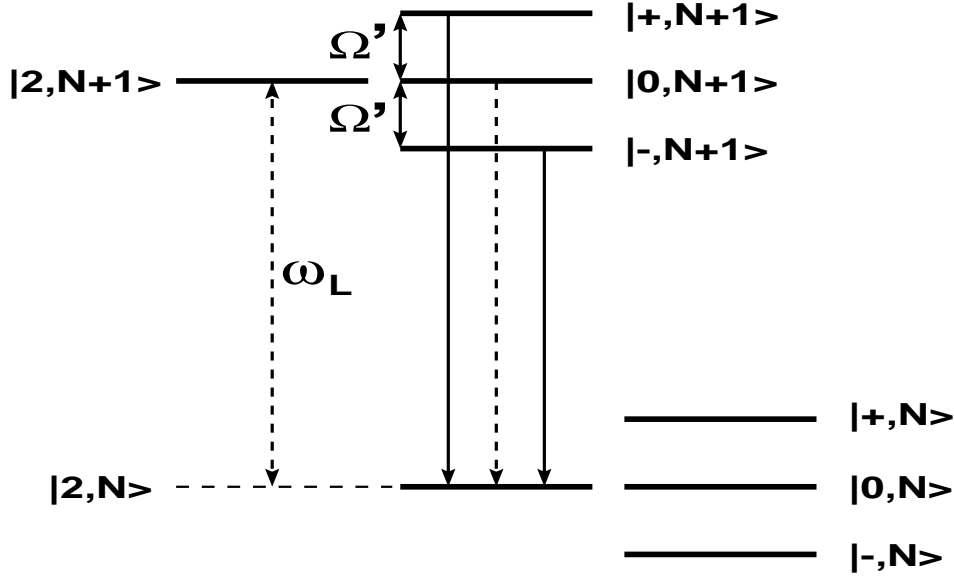}
\caption{Dressed states of two neighboring manifolds, $N+1$
and $N$. Solid arrows indicate transitions at $\omega_{L}\pm \Omega^{\prime}$ which are only slightly affected by quantum interference, while the dashed arrow indicates the transition at the laser frequency $\omega_{L}$ which is strongly affected
by quantum interference and vanishes for parallel dipole moments and $|\mbox{\boldmath $\mu$}_{12}|=|\mbox{\boldmath $\mu$}_{32}|$.}
\end{center}
\end{figure}

Dressed states of two neighbouring manifolds are shown in Fig.~4. The
manifolds are separated by $\omega _{L}$, while the states inside each
manifold are separated by $\Omega ^{\prime }$. The dressed states are
connected by transition dipole moments. It is easily verified that non-zero
dipole moments occur only between states within neighbouring manifolds.
Defining transition dipole moments $\mbox{\boldmath $\mu$}%
_{i,N+1;2,N}=\left\langle N+1,i\right| \mbox{\boldmath $\mu$}\left|
2,N\right\rangle $ between $\left| i,N+1\right\rangle $ $(i=0,-,+)$ and $%
\left| 2,N\right\rangle ,$ and using Eq.~(\ref{75}), we find 
\begin{eqnarray}
\mbox{\boldmath $\mu$}_{+,N+1;2,N} &=&\left( 1-\alpha \right) 
\mbox{\boldmath
$\mu$}_{12}+\left( 1+\alpha \right) \mbox{\boldmath $\mu$}_{32},  \nonumber
\\
\mbox{\boldmath $\mu$}_{0,N+1;2,N} &=&-\beta \left( \mbox{\boldmath $\mu$}%
_{12}-\mbox{\boldmath $\mu$}_{32}\right) ,  \nonumber \\
\mbox{\boldmath $\mu$}_{-,N+1;2,N} &=&-\left[ \left( 1+\alpha \right) %
\mbox{\boldmath $\mu$}_{12}+\left( 1-\alpha \right) \mbox{\boldmath $\mu$}%
_{32}\right] .  \label{77}
\end{eqnarray}
The transition dipole moments $\mbox{\boldmath $\mu$}_{2,N;i,N-1}$ between $%
\left| 2,N\right\rangle $ and the dressed states $\left| i,N-1\right\rangle $
of the manifold below are equal to zero, independent of the mutual
orientation of the atomic dipole moments. It is evident from Eq.~(\ref{77})
that transitions to the state $\left| 2,N\right\rangle $ depend on the
mutual polarization of the dipole moments $\mbox{\boldmath $\mu$}_{12}$ and $%
\mbox{\boldmath $\mu$}_{32}$. For $\mbox{\boldmath $\mu$}_{12}\parallel %
\mbox{\boldmath $\mu$}_{32}$ and $|\mbox{\boldmath $\mu$}_{12}|=|%
\mbox{\boldmath $\mu$}_{32}|$ the transition dipole moment $%
\mbox{\boldmath
$\mu$}_{0,N+1;2,N}$ vanishes, resulting in the disappearance of the central
component of the fluorescence spectrum. When $\mbox{\boldmath $\mu$}_{12}$
and $\mbox{\boldmath $\mu$}_{32}$ are not parallel, all the transitions are
allowed, and three lines can be seen in the spectrum.

It is interesting to note that in the case of antiparallel dipole moments
and $\Delta =0$ the dipole moments $\mbox{\boldmath
$\mu$}_{+,N+1;2,N}$ and $\mbox{\boldmath $\mu$}_{-,N+1;2,N}$ vanish,
resulting in the disappearance of the Rabi sidebands of the spectrum. Thus,
depending on the polarization of the dipole moments and the splitting $%
\Delta $, the spectrum can exhibit, one, two or three spectral lines. The
dressed-atom model clearly explains the origin of the cancellation of the
spectral lines. This effect arises from the cancellation of the transition
dipole moments due to quantum interference between the two atomic
transitions.

\subsection{Excitation of a single transition}

\label{sec7b}

Here, we consider a three-level $V$-type atom driven by a strong laser field
of Rabi frequency $\Omega $, coupled {\it solely} to the $\left|
1\right\rangle -\left| 2\right\rangle $ transition. This is a crucial
assumption, which would be difficult to realize in practice since quantum
interference requires almost parallel dipole moments. However, the
difficulty can be overcome in atomic systems with specific selection rules
for the transition dipole moments, or by applying fields with specific
polarization properties~\cite{ma98}.

In addition, we assume that the atom is probed by a weak laser field. We
consider two different coupling configurations of the probe beam to the
driven atom. In the first case, we assume that the probe beam is exclusively
coupled to the driven $\left| 1\right\rangle \rightarrow \left|
2\right\rangle $ transition~\cite{afsjmo}. In the second case, we will
assume that the probe beam is coupled to the undriven $\left| 3\right\rangle
\rightarrow \left| 2\right\rangle $ transition. For the second case, Menon
and Agarwal~\cite{ma00} have predicted that in the presence of quantum
interference the absorption spectrum of the probe beam can exhibit gain
features instead of the usual Autler-Townes doublet. This unexpected feature
requires the condition that the driving field couples to only one of the two
atomic transitions and the Rabi frequency $\Omega $ of the driving field is
such that $\Omega =2\Delta$, where $\Delta$ is the splitting between the
excited states.

The absorption rate of a probe beam of a tunable frequency $\omega_{p}$
monitoring the $\left| 1\right\rangle -\left| 2\right\rangle $ transition is
defined as~\cite{ma00,mol72} 
\begin{equation}
W_{12}\left(\omega_{p}\right) = {\rm Re}\left[\Omega_{p}
\rho_{12}^{\left(+1\right)}\right] ,  \label{78}
\end{equation}
where $\Omega_{p}$ is the Rabi frequency of the probe beam, and $%
\rho_{12}^{\left(+1\right)}$ is the stationary component (harmonic) of the
coherence $\rho_{12}$ oscillating with the probe detuning $\delta
=\omega_{p}-\omega_{2}$.

\begin{figure}
\begin{center}
\includegraphics[height=3.5in,width=2.5in]{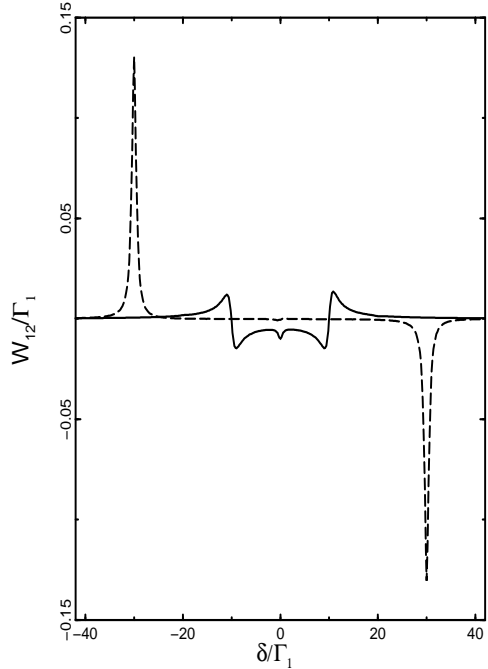}
\caption{The absorption rate $W_{12}$ as a function of
$\delta/\Gamma_{1}$ for $p=0.95, \Gamma_{1}=\Gamma_{2},
\Omega_{p}=0.5\Gamma_{1},
\Delta=15\Gamma_{1}$ and different $\Omega$: $\Omega=10\Gamma_{1}$
(solid line) and $\Omega=30\Gamma_{1}$ (dashed line).}
\end{center}
\end{figure}

In Fig.~5, we plot the absorption rate $W_{12}$ as a function of $\delta $
for $p=0.95$ and different $\Omega $. When $\Omega \neq 2\Delta $ the
absorption rate exhibits the familiar Mollow absorption spectrum~\cite{mol72}
with small dispersive structures at $\delta =\pm \Omega $. The absorption
rate changes dramatically when $\Omega =2\Delta $. Here, the dominant
features of the rate are emissive and absorptive components at $\delta =\pm
\Omega $, indicating that at $\delta =-\Omega $ the weaker field is
absorbed, whereas at $\delta =\Omega $ it is amplified at the expense of the
strong field. The weaker field is always absorbed (amplified) at $\delta
=-\Omega $ $(\delta =\Omega )$ independent of the ratio $r=\Gamma
_{1}/\Gamma _{2}$ between the spontaneous emission rates $\Gamma _{1}$ and $%
\Gamma _{2}$. Note that the absorption rate shown in Fig.~5 is similar to
the Mollow absorption spectrum for an off-resonant driving field~\cite{mol72}%
. However, there is a significant difference in that the ratio between the
magnitudes of the emissive and absorptive peaks in the Mollow spectrum is
always less than one and the ratio varies with the detuning and Rabi
frequency of the driving field. The ratio of the absorption rates, shown in
Fig.~5, is equal to one and constant independent of the values of the
parameters involved.

\begin{figure}
\begin{center}
\includegraphics[height=3.5in,width=3in]{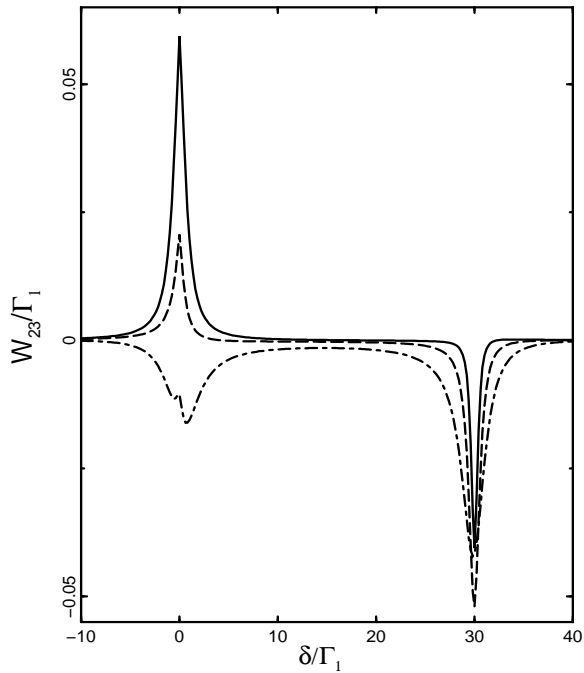}
\caption{The absorption rate $W_{23}$ as a function of $\delta$ for
$\Omega=30\Gamma_{1}$, $\Omega_{p}=0.5\Gamma_{1}$, $\Delta=15\Gamma_{1}$, $p=0.95$ and different values of $r$: $r=1$ (solid line), $r=2$ (dashed line) and $r=5$ (dashed-dotted line).}
\end{center}
\end{figure}

In Fig.~6, we present the absorption rate for the case considered by Menon
and Agarwal~\cite{ma00}, in which the probe beam is coupled to the undriven $%
|3\rangle -|2\rangle $ transition 
\begin{equation}
W_{23}\left( \omega _{p}\right) =2{\rm Re}\left[ \Omega _{p}\rho
_{23}^{\left( +1\right) }\right] .  \label{79}
\end{equation}
The absorption rate is plotted as a function of $\delta $ for $\Omega
=2\Delta $. We see that the absorption rate exhibits an emissive feature at $%
\delta =\Omega $. Moreover, there is a central component at $\delta =0$,
whose absorptive/emissive properties depend on the ratio $r$. For $r<2$ the
rate is positive, indicating that the weaker field is absorbed by the
system. As $r$ increases the absorptive feature decreases and vanishes for $%
r\approx 2$. When we further increase $r$ $(r>2)$ the absorptive features at 
$\delta =0$ switch into emissive features and the magnitude of the emissive
peak increases with increasing $r$. The threshold value for $r$, at which
absorption switches to emission, depends on $p$. For $p=1$ the threshold is
exactly at $r=2$, and shifts towards larger $r$ as $p$ decreases.

The physics associated with the unusual properties of the absorption rate of
the probe beam, shown in Figs.~5 and 6, can be easily explored by working in
the basis of quantum dressed states of the system~\cite{fsa01}. In the case
of the driving laser coupled exclusively to the $|1\rangle -|2\rangle $
transition, the Hamiltonian of the system can be written as 
\begin{equation}
H_{S}=H_{0}+H_{int},  \label{80}
\end{equation}
where 
\begin{equation}
H_{0}=\hbar \omega _{1}S_{1}^{+}S_{1}^{-}+\hbar \omega
_{2}S_{2}^{+}S_{2}^{-}+\hbar \omega _{1}a_{L_{1}}^{\dagger }a_{L_{1}},
\label{81}
\end{equation}
is the Hamiltonian of the atom plus driving field, and 
\begin{equation}
H_{int}=-\frac{1}{2}\hbar g\left( a_{L_{1}}^{\dagger
}S_{1}^{-}+a_{L_{1}}S_{1}^{+}\right)   \label{82}
\end{equation}
is the interaction between the atom and the laser field.

The Hamiltonian $H_{0}$ has the ``undressed'' eigenstates $|1,N-1\rangle,
|3,N-1\rangle$ and $|2,N\rangle$. The states $|1,N-1\rangle$ and $%
|2,N\rangle $ are degenerate with energies $E_{1,N}=E_{2,N}=\hbar\omega_{1}$%
, while the state $|3,N-1\rangle$ has energy $E_{3,N}=\hbar
(N\omega_{1}+\Delta)$, where $N$ is the number of photons in the laser mode.
When we include the interaction~(\ref{82}) between the atom and the laser
field, the degeneracy is lifted, resulting in triplets of dressed states 
\begin{eqnarray}
|+,N\rangle &=& \frac{1}{\sqrt{2}}\left(|2,N\rangle+|1,N-1\rangle\right), 
\nonumber \\
|-,N\rangle &=& \frac{1}{\sqrt{2}}\left(|2,N\rangle-|1,N-1\rangle\right), 
\nonumber \\
|\tilde{3},N\rangle &=& |3,N-1\rangle ,  \label{83}
\end{eqnarray}
with energies 
\begin{eqnarray}
E_{+,N} &=& \hbar \left(N\omega_{1} +\frac{1}{2}\Omega \right) ,  \nonumber
\\
E_{-,N} &=& \hbar \left(N\omega_{1} -\frac{1}{2}\Omega \right) ,  \nonumber
\\
E_{\tilde{3},N} &=& \hbar \left(N\omega_{1} +\Delta \right) .  \label{84}
\end{eqnarray}
The dressed states~(\ref{83}) group into manifolds of nondegenerate triplets
unless $\Delta =\frac{1}{2}\Omega$ and then the states $|+,N\rangle$, $|%
\tilde{3},N\rangle$ in each manifold are degenerate.

Since the driven and undriven transitions are coupled through the $\Gamma
_{12}$ terms, it is convenient to introduce symmetric and anti-symmetric
superposition states of the dressed states $|+,N\rangle $ and $|\tilde{3}%
,N\rangle $. According to Eq.~(\ref{84}), the superposition states
diagonalise the dissipative (damping) part of the master equation of the
system. The superposition states can be written as~\cite{afsjmo} 
\begin{eqnarray}
|s,N\rangle  &=&\alpha |+,N\rangle +\beta |\tilde{3},N\rangle ,  \label{85}
\\
|a,N\rangle  &=&\beta |+,N\rangle -\alpha |\tilde{3},N\rangle ,  \label{86}
\end{eqnarray}
where 
\begin{equation}
\alpha =\frac{1}{\sqrt{1+2r}},\beta =\frac{\sqrt{2r}}{\sqrt{1+2r}},
\label{87}
\end{equation}
and $r=\Gamma _{2}/\Gamma _{1}$.

With the dressed states of the driven system available, we may easily
predict transition frequencies and calculate transition dipole moments and
spontaneous emission rates between the dressed states of the system. It is
easily verified that non-zero dipole moments occur only between dressed
states within neighbouring manifolds. Using Eqs.~(\ref{85}) and (\ref{86}),
we find that the transition dipole moments between $|N,i\rangle $ and $%
|N-1,j\rangle $ are 
\begin{eqnarray}
\mbox{\boldmath $\mu$}_{s,N;s,N-1} &=&\frac{1}{2}\alpha \left( \alpha %
\mbox{\boldmath $\mu$}_{12}+\sqrt{2}\beta \mbox{\boldmath $\mu$}_{32}\right)
,\quad \mbox{\boldmath $\mu$}_{s,N;-,N-1}=\frac{1}{2}\left( \alpha %
\mbox{\boldmath $\mu$}_{12}+\sqrt{2}\beta \mbox{\boldmath $\mu$}_{32}\right)
,  \nonumber \\
\mbox{\boldmath $\mu$}_{-,N;s,N-1} &=&-\frac{1}{2}\alpha 
\mbox{\boldmath
$\mu$}_{12},\quad \mbox{\boldmath $\mu$}_{-,N;-,N-1}=-\frac{1}{2}%
\mbox{\boldmath $\mu$}_{12},  \nonumber \\
\mbox{\boldmath $\mu$}_{-,N;a,N-1} &=&-\frac{1}{2}\beta 
\mbox{\boldmath
$\mu$}_{12},\quad \mbox{\boldmath $\mu$}_{s,N;a,N-1}=\frac{1}{2}\beta \left(
\alpha \mbox{\boldmath $\mu$}_{12}+\sqrt{2}\beta \mbox{\boldmath $\mu$}%
_{32}\right) ,  \nonumber \\
\mbox{\boldmath $\mu$}_{a,N;s,N-1} &=&\frac{1}{2}\alpha \left( \beta %
\mbox{\boldmath $\mu$}_{12}-\sqrt{2}\alpha \mbox{\boldmath $\mu$}%
_{32}\right) ,\quad \mbox{\boldmath $\mu$}_{a,N;-,N-1}=\frac{1}{2}\left(
\beta \mbox{\boldmath $\mu$}_{12}-\sqrt{2}\alpha \mbox{\boldmath $\mu$}%
_{32}\right) ,  \nonumber \\
\mbox{\boldmath $\mu$}_{a,N;a,N-1} &=&\frac{1}{2}\beta \left( \beta %
\mbox{\boldmath $\mu$}_{12}-\sqrt{2}\alpha \mbox{\boldmath $\mu$}%
_{32}\right) ,  \label{88}
\end{eqnarray}
where $\mbox{\boldmath $\mu$}_{i,N;j,N-1}=\langle i,N|\tilde{%
\mbox{\boldmath
$\mu$}}|j,N-1\rangle $, and $\tilde{\mbox{\boldmath $\mu$}}=\tilde{%
\mbox{\boldmath $\mu$}}_{1}+\tilde{\mbox{\boldmath $\mu$}}_{2}$ is the total
dipole moment of the atom.

The spontaneous transitions occur with probabilities $\Gamma _{i,N;j,N-1}$
given by 
\begin{equation}
\Gamma _{i,N;j,N-1}=\frac{\Gamma_{n}}{\left|\mbox{\boldmath $\mu$}_{2n-1,2}
\right|^{2}}\left|\left\langle i,N\right|\tilde{\mbox{\boldmath $\mu$}}
\left| j,N-1\right\rangle \right|^{2},\quad n=1,2 ,  \label{89}
\end{equation}

\begin{figure}
\begin{center}
\includegraphics[height=3in,width=4in]{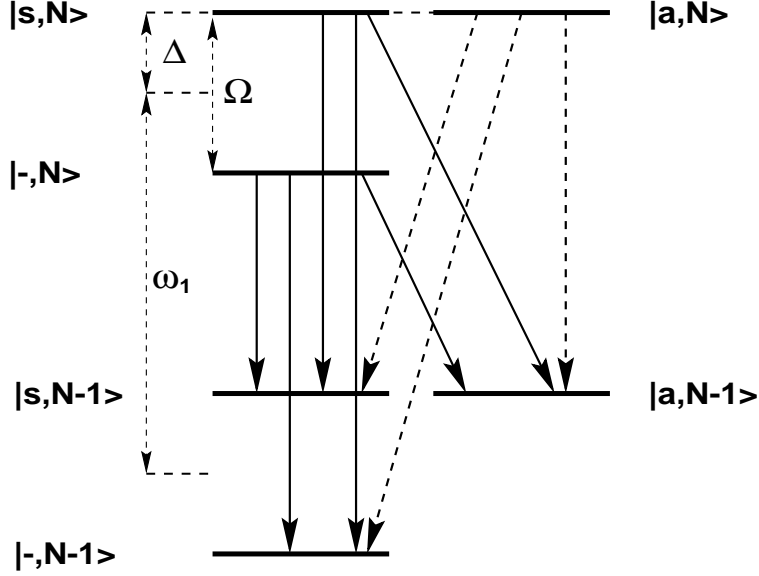}
\caption{Energy level diagram of the superposition dressed states
for $\Delta =\frac{1}{2}\Omega$. The solid lines indicate spontaneous transitions which occur independent of quantum interference, whereas the dash ed lines indicate transitions which are significantly reduced by quantum interference.}
\end{center}
\end{figure}

In Fig.~7, we present the dressed states of two neighbouring manifolds, $N$
and $N-1$, and the possible transitions among them. Solid lines indicate
transitions which are not significantly affected by quantum interference,
whereas dashed lines indicate transitions which are strongly modified by
quantum interference, in that their transition dipole moments decrease with
increasing $p$ and vanish for $p=1$. We see from Fig.~7 that quantum
interference strongly affects transition rates from the antisymmetric state
to the states of the manifold below. Thus, the antisymmetric state becomes a
dark state in the limit of strong interference, $p\approx 1$. Moreover, it
follows from the master equation~(\ref{29}) and the dressed state~(\ref{86})
that in the steady-state the antisymmetric state is strongly populated, and
the population is trapped in the antisymmetric state $(P_{a}=1)$ when $%
\theta =0$.

Figure~7, together with the transition dipole moments and transition rates,
provides a simple interpretation of the absorption rate shown in Fig.~5.
Since the antisymmetric state is strongly populated, the emissive peak in
the absorption rate appears on an almost completely inverted transition $%
(|a,N\rangle -|-,N-1\rangle )$, whose dipole moment is significantly reduced
by quantum interference. One might expect that the weaker field should not
couple to an almost canceled dipole moment. However, we have assumed that
the probe field couples {\it only} to the dipole moment $\mbox{\boldmath
$\mu$}_{12}$. From Eq.~(\ref{88}), we find that the coupling strength of the
probe field to the transition $|a,N\rangle -|-,N-1\rangle $ is proportional
to $\frac{1}{2}\beta \mbox{\boldmath $\mu$}_{12}$ despite the fact that the
total dipole moment of the transition is much smaller, $\mbox{\boldmath
$\mu$}_{a,N;-,N-1}=\frac{1}{2}\beta \mbox{\boldmath$\mu$}_{12}-\frac{1}{%
\sqrt{2}}\alpha \mbox{\boldmath $\mu$}_{32}$. The absorptive peak, seen in
Fig.~5 at the frequency $\omega _{1}-\Omega $, appears on the non-inverted
transition $|-,N\rangle -|a,N-1\rangle $ with the transition dipole moment $%
\frac{1}{2}\beta \mbox{\boldmath $\mu$}_{12}$. Since the absolute values of
the population difference on the $|a,N\rangle -|-,N-1\rangle $ and $%
|-,N\rangle -|a,N-1\rangle $ transitions are the same and the coupling
strengths of the weaker field to the transitions are equal, $\mbox{\boldmath
$\mu$}_{a,N;-,N-1}=\mbox{\boldmath $\mu$}_{-,N;a,N-1}=\frac{1}{2}\beta %
\mbox{\boldmath $\mu$}_{12}$, the absolute values of the absorptive and
emissive peaks in the absorption rate are the same, independent of the ratio 
$r=\Gamma _{1}/\Gamma _{2}$. One sees from Fig.~7 that there are two
transitions, one emissive $(|a,N\rangle -|s,N-1\rangle )$ and one absorptive 
$(|s,N\rangle -|a,N-1\rangle )$, which contribute to the central structure
at $\delta =0$. Since the absolute values of the population difference on
these transitions are the same and the coupling strengths of the weaker
field to these transitions are equal, $\mbox{\boldmath $\mu$}_{s,N;a,N-1}=%
\mbox{\boldmath $\mu$}_{a,N;s,N-1}=\frac{1}{2}\alpha \beta 
\mbox{\boldmath
$\mu$}_{12}$, these two contributions cancel each other leading to a
transparency of the weaker field at $\delta =0$.

The physical origin of the gain features shown in Fig.~6 can also be
explained with the help of the energy-level diagram of Fig.~7 and the
transition dipole moments~(\ref{88}). Since the weaker field couples
exclusively to $\mbox{\boldmath $\mu$}_{32}$, the transition $\left|
a,N-1\right\rangle -\left| -,N\right\rangle $, whose dipole moment is
proportional to $\mbox{\boldmath $\mu$}_{12}$, is transparent for the weaker
field. The coupling strength of the weaker field to the $\left|
a,N\right\rangle -\left| -,N-1\right\rangle $ transition is proportional to $%
{\displaystyle{1 \over \sqrt{2}}}%
\alpha \mbox{\boldmath$\mu$}_{32}$ indicating that the field can be
amplified on this transition and the amplification is not much affected by
the the ratio $r$. It is seen from Fig.~7 that at $\delta =0$ the probe
couples to three transitions. The transition $\left| a,N\right\rangle
-\left| a,N-1\right\rangle $ is transparent for the probe because it occurs
between two states of the same population. Therefore, the
absorptive/emissive properties result from the coupling of the probe to the $%
\left| s,N\right\rangle -\left| a,N-1\right\rangle $ and $\left|
a,N\right\rangle -\left| s,N-1\right\rangle $ transitions. For $\theta
\approx 0$ almost all the population is trapped in the antisymmetric state,
and then the probe is strongly absorbed on the $\left| s,N\right\rangle
-\left| a,N-1\right\rangle $ transition, but is amplified on the $\left|
a,N\right\rangle -\left| s,N-1\right\rangle $ transition. According to Eq.~(%
\ref{88}), the latter is a dark transition. Since the absolute values of the
population difference between the states are the same for both transitions,
the absorptive/emissive properties at $\delta =0$ depend solely on the
relation between the transition rates. From Eq.~(\ref{88}), we find that the
coupling strength of the probe beam to the transition $\left|
a,N\right\rangle -\left| s,N-1\right\rangle $ is proportional to $\frac{1}{%
\sqrt{2}}\alpha ^{2}\mbox{\boldmath $\mu$}_{32}$, whereas the coupling
strength to the transition $\left| s,N\right\rangle -\left|
a,N-1\right\rangle $ is proportional to $\frac{1}{\sqrt{2}}\beta ^{2}%
\mbox{\boldmath $\mu$}_{32}$. Thus, the absorptive/emissive properties at $%
\delta =0$ depend on the difference $(\beta ^{2}-\alpha ^{2})=\frac{1}{2}%
\beta ^{2}(2-r)$. For $r<2$ the difference is positive, indicating that the
weaker field is absorbed at $\delta =0$, and is amplified for $r>2$. These
simple dressed atom predictions are in excellent agreement with the
numerical calculations shown in Fig.~6.

Thus, in terms of the quantum dressed-states the gain features predicted by
Menon and Agarwal~\cite{ma00} actually appear on completely inverted
transitions whose dipole moments are canceled by quantum interference.
Therefore, the gain features can be regarded as amplifications on dark
transitions~\cite{fsa01}.

\subsection{Both transitions excited}

\label{sec7c}

Another aspect of quantum interference effects which has been studied
extensively, is the response of a $V$-type three-level atom to a coherent
laser field directly coupled to the decaying transitions. This was studied
by Cardimona {\it et al.}~\cite{crs}, who found that the system can be
driven into a trapping state in which quantum interference prevents any
fluorescence from the excited levels, regardless of the intensity of the
driving laser. Similar predictions have been reported by Zhou and Swain~\cite
{zs96}, who have shown that ultrasharp spectral lines can be predicted in
the fluorescence spectrum when the dipole moments of the atomic transitions
are nearly parallel and the fluorescence can be completely quenched when the
dipole moments are exactly parallel.

When the atomic transitions $|{1}\rangle\rightarrow |{2}\rangle$ and $|{3}%
\rangle\rightarrow |{2}\rangle$ are directly driven by a laser field, the
master equation~(\ref{29}) leads to the following set of equations of
motions for the density matrix elements 
\begin{eqnarray}
\dot{\tilde{\rho}}_{12} &=&(\dot{\tilde{\rho}}_{21})^{\ast }=\frac{1}{2}i
\Omega_{1} -\left[ \frac{1}{2}\Gamma_{1}-i\left(\Delta_{L}-\frac{1}{2}%
\Delta\right)\right] \tilde{\rho}_{12} -\frac{1}{2}\Gamma_{12}\tilde{\rho}%
_{32} - \frac{1}{2}i\Omega_{2}\rho_{13}-\frac{1}{2}i\Omega_{1}(2\rho_{11}
+\rho_{33}) ,  \nonumber \\
\dot{\tilde{\rho}}_{32} &=&(\dot{\tilde{\rho}}_{23})^{\ast }=\frac{1}{2}i
\Omega_{2} -\left[\frac{1}{2}\Gamma_{2}-i\left(\Delta_{L}+\frac{1}{2}%
\Delta\right) \right]\tilde{\rho}_{32} -\frac{1}{2}\Gamma_{12}\tilde{\rho}%
_{12} - \frac{1}{2}i\Omega_{1}\rho_{31}-\frac{1}{2}i\Omega_{2}
(2\rho_{33}+\rho_{11}) ,  \nonumber \\
\dot{\rho}_{31} &=&(\dot{\rho}_{13})^{\ast }= -\left[\frac{1}{2}%
(\Gamma_{1}+\Gamma_{2})-i\Delta \right]\rho_{31} -\frac{1}{2}%
\Gamma_{12}(\rho _{33}+\rho _{11}) - \frac{1}{2}i\Omega_{1}\tilde{\rho}_{32}+%
\frac{1}{2}i\Omega_{2} \tilde{\rho}_{21} ,  \nonumber \\
\dot{\rho}_{11} &=&-\Gamma_{1}\rho_{11}-\frac{1}{2}\Gamma_{12}(\rho_{13}
+\rho_{31})+\frac{1}{2}i\Omega_{1}(\tilde{\rho}_{21} -\tilde{\rho}_{12}) , 
\nonumber \\
\dot{\rho}_{33} &=&-\Gamma_{2}\rho_{33}-\frac{1}{2}\Gamma_{12}(\rho_{13}
+\rho_{31})+\frac{1}{2}i\Omega_{2}(\tilde{\rho}_{23} -\tilde{\rho}_{32}) ,
\label{90}
\end{eqnarray}
where 
\begin{equation}
\tilde{\rho}_{j2} = \rho_{j2}\exp \left[i\left(\omega_{L}t+\phi_{L}\right)%
\right] ,\quad (j=1,3) ,  \label{91}
\end{equation}
and $\Delta_{L} =\omega_{L}-\frac{1}{2}\left(\omega_{1}+\omega_{2}\right)$
is the detuning of the laser frequency from the middle of the upper levels
splitting.

\begin{figure}
\begin{center}
\includegraphics[height=3.5in,width=4.5in]{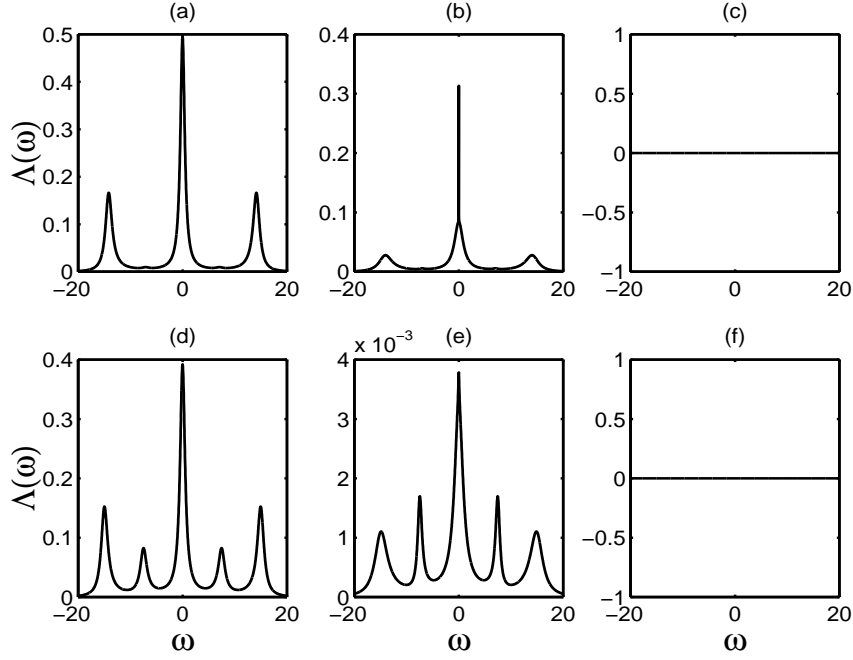}
\caption{The fluorescence spectrum for the $V$-type
three-level atom with
non-degenerate transitions driven by a strong laser field of the Rabi
frequency $\Omega =5\Gamma_{1}$, $\Delta_{L}=0, \Gamma_{1}=\Gamma_{2}=\Gamma$
and different $\Delta$ and $\Gamma_{12}$: (a) $\Delta =\Gamma, \Gamma_{12}=0$,
(b) $\Delta =\Gamma, \Gamma_{12}= 0.999\Gamma$, (c)
$\Delta =\Gamma, \Gamma_{12}= \Gamma$, (d)
$\Delta =5\Gamma, \Gamma_{12}= 0$, (e)
$\Delta =5\Gamma, \Gamma_{12}= 0.999\Gamma$, (f)
$\Delta =5\Gamma, \Gamma_{12}= \Gamma$.}
\end{center}
\end{figure}

We apply the equations of motion~(\ref{90}) to calculate numerically the
steady-state fluorescence spectrum of the driven atom. In Fig.~8, we plot
the fluorescence spectrum for a strong driving field tuned to the middle of
the upper levels splitting $\Delta $. For small $\Delta $ the spectrum
exhibits a three-peak structure, similar to the Mollow spectrum of a
two-level atom~\cite{mol69}, while for large $\Delta $ the spectrum consists
of five peaks whose intensities and widths vary with the cross-damping term $%
\Gamma _{12}$. When the dipole moments are nearly parallel, $\Gamma
_{12}=0.999\Gamma $, a significant sharp peak appears at the central
frequency superimposed on a broad peak. However, in the case of exactly
parallel dipole moments, $\Gamma _{12}=\Gamma $, and the fluorescence
emission quenches completely at all frequencies.

The dependence of the number of peaks on the splitting $\Delta $, and the
variation of their intensities and widths with $\Gamma _{12}$ can be readily
explained in terms of transition rates between dressed states of the system.
For the three-level system discussed here, the Hamiltonian is given by 
\begin{equation}
H^{\prime }=H_{0}+V_{L},  \label{92}
\end{equation}
where 
\begin{equation}
H_{0}=-\hbar \left( \Delta _{L}-\frac{1}{2}\Delta \right)
S_{1}^{+}S_{1}^{-}-\hbar \left( \Delta _{L}+\frac{1}{2}\Delta \right)
S_{2}^{+}S_{2}^{-}+\hbar \omega _{L}a_{L}^{\dagger }a_{L}  \label{93}
\end{equation}
is the Hamiltonian of the uncoupled system, and 
\begin{equation}
V_{L}=-\frac{\hbar }{2}g\left[ a_{L}^{\dagger }\left(
S_{1}^{-}+S_{2}^{-}\right) +\left( S_{1}^{+}+S_{2}^{+}\right) a_{L}\right] 
\label{94}
\end{equation}
is the interaction between the laser field and the atomic transitions.

For $\Delta _{L}=0$ the Hamiltonian $H_{0}$ has three non-degenerate
eigenstates $\left| 2,N\right\rangle,$ $\left| 1,N-1\right\rangle $, and $%
\left|3,N-1\right\rangle $, where $\left| i,N\right\rangle $ is the state
with the atom in state $\left| i\right\rangle $ and $N$ photons present in
the driving laser mode. When we include the interaction $V_{L}$ the triplets
recombine into new triplets with eigenvectors (dressed states) 
\begin{eqnarray}
\left| +,N\right\rangle &=&\frac{1}{2}\left[\left(1-\alpha\right) \left|
1,N-1\right\rangle + \left(1+\alpha\right) \left| 3,N-1\right\rangle -
2\beta \left|2,N\right\rangle \right] ,  \nonumber \\
\left| 0,N\right\rangle &=& -\beta \left(\left|1,N-1\right\rangle
-\left|3,N-1\right\rangle\right) +\alpha \left| 2,N\right\rangle ,
\label{95} \\
\left| -,N\right\rangle &=& \frac{-1}{2}\left[\left(1+\alpha\right) \left|
1,N-1\right\rangle + \left(1-\alpha\right) \left| 3,N-1\right\rangle +
2\beta \left| 2,N\right\rangle \right] ,  \nonumber
\end{eqnarray}
corresponding to energies 
\begin{eqnarray}
E_{N,+} &=&\hbar \left(N\omega_{L}+\tilde{\Omega}\right) ,  \nonumber \\
E_{N,0} &=&\hbar N\omega _{L},  \nonumber \\
E_{N,-} &=&\hbar \left(N\omega _{L}-\tilde{\Omega}\right) ,  \label{96}
\end{eqnarray}
where $\tilde{\Omega} =\sqrt{\Delta ^{2}+\frac{1}{2}\Omega ^{2}},\alpha
=\Delta /2\tilde{\Omega}$ and $\beta =\Omega /2\tilde{\Omega}$.

The dressed states (\ref{95}) group into manifolds, each containing three
states. Neighbouring manifolds are separated by $\omega _{L}$, while the
states inside each manifold are separated by $\tilde{\Omega}$. Interaction
between the atom and the vacuum field leads to a spontaneous emission
cascade down its energy manifold ladder. The probability of a transition
between any two dressed states is proportional to the absolute square of the
the dipole transition moment between these states. It is easily verified
that non-zero dipole moments occur only between states within neighbouring
manifolds. Using (\ref{95}) and assuming that $\mbox{\boldmath
$\mu$}_{13}=\mbox{\boldmath $\mu$}_{23}=\mbox{\boldmath $\mu$}$, we find
that the transition dipole moments $\mbox{\boldmath
$\mu$}_{i,N+1;0,N}=\left\langle N+1,i\right| \mbox{\boldmath
$\mu$}\left| 0,N\right\rangle $ between $\left| 0,N\right\rangle $ and the
dressed states of the manifold above are 
\begin{eqnarray}
\mbox{\boldmath $\mu$}_{+,N+1;0,N} &=&\frac{1}{2}\alpha 
\mbox{\boldmath
$\mu$}\left[ \left( 1-\alpha \right) +\left( 1+\alpha \right) \cos \theta %
\right] ,  \nonumber \\
\mbox{\boldmath $\mu$}_{0,N+1;0,N} &=&-\alpha \beta \mbox{\boldmath $\mu$}%
\left( 1-\cos \theta \right) ,  \nonumber \\
\mbox{\boldmath $\mu$}_{-,N+1;0,N} &=&-\frac{1}{2}\alpha 
\mbox{\boldmath
$\mu$}\left[ \left( 1+\alpha \right) +\left( 1-\alpha \right) \cos \theta %
\right] ,  \label{97}
\end{eqnarray}
whereas the transition dipole moments between $\left| 0,N\right\rangle $ and
the dressed states of the manifold below are 
\begin{eqnarray}
\mbox{\boldmath $\mu$}_{0,N;+,N-1} &=&\beta ^{2}\mbox{\boldmath $\mu$}\left(
1-\cos \theta \right) ,  \nonumber \\
\mbox{\boldmath $\mu$}_{0,N;0,N-1} &=&-\alpha \beta \mbox{\boldmath $\mu$}%
\left( 1-\cos \theta \right) ,  \nonumber \\
\mbox{\boldmath $\mu$}_{0,N;-,N-1} &=&\beta ^{2}\mbox{\boldmath $\mu$}\left(
1-\cos \theta \right) ,  \label{98}
\end{eqnarray}
where $\theta $ is the angle between the dipole moments.

It is apparent from Eq.~(\ref{98}) that transitions from the state $\left|
0,N\right\rangle $ to the dressed states of the manifold below are allowed
only if the dipole moments are not parallel. The transitions occur with
significantly reduced rates, proportional to $(1-\cos \theta )$, giving very
narrow lines when $\theta \approx 0^{o}$. For parallel dipole moments, the
transitions to the state $\left| 0,N\right\rangle $ are allowed from the
dressed states of the manifold above, but are forbidden to the states of the
manifold below. Therefore, the state $\left| 0,N\right\rangle $ is a
trapping state such that the population can flow into this state but cannot
leave it, resulting in the disappearance of the fluorescence from the driven
atom. The non-zero transition rates to the state $\left| 0,N\right\rangle $
are proportional to $\alpha $ and are allowed only when $\Delta \neq 0$.
Otherwise, for $\Delta =0$, the state $\left| 0,N\right\rangle $ is
completely decoupled from the remaining dressed states. In this case the
three-level system is equivalent to a two-level atom.

The above dressed-atom analysis shows that quantum interference and the
driving laser field create a ``dressed'' trapping state which is a linear
superposition of the $|{a}\rangle $ and $|{2}\rangle $ states. This trapping
state is different from the trapping state created by quantum interference
in the absence of the driving field, see Eq.~(\ref{63}), which is the
antisymmetric state $|{a}\rangle $ alone. As seen from Eq.~(\ref{95}), the
dressed trapping state reduces to the state $|{a}\rangle $ for a very strong
driving field $(\Omega \gg \Delta )$.

The narrow resonances produced by quantum interference may also be observed
in the absorption spectrum of a three-level atom probed by a weak field of
the frequency $\omega_{p}$. Zhou and Swain~\cite{zs97} have calculated the
absorption spectrum of a probe field monitoring $V$-type three-level atoms
with degenerate $(\Delta =0)$ as well as non-degenerate $(\Delta \neq 0)$
transitions and have demonstrated that quantum interference between the two
atomic transitions can result in very narrow spectral lines, transparency,
and even gain without population inversion. Paspalakis {\it et al.}~\cite
{pkoc} have calculated the absorption spectrum and refractive index of a $V$%
-type three-level atom driven by coherent and incoherent fields and have
found that quantum interference enhances the index of refraction and can
produce a very strong gain without population inversion.

\subsection{Three-level $\Lambda$ system}

\label{sec7d}

It has been known for a long time, that in a $\Lambda $-type three-level
atom with two transitions with perpendicular dipole moments $(\Gamma _{12}=0)
$ driven by two laser fields, the population can be trapped in the ground
states of the atom. This phenomenon, known as coherent population trapping
(CPT) has been theoretically investigated by Arimondo and Orriols~\cite{ao},
Gray {\it et al.}~\cite{gray}, Orriols~\cite{orr}, and experimentally
observed by Alzeta {\it et al.}~\cite{alz}. Coherent population trapping has
been examined in review articles by Dalton and Knight~\cite{dk} and
Arimondo~ \cite{ari}. Javanainen~\cite{java}, Ferguson {\it et al.}~\cite
{ffd} and Menon and Agarwal~\cite{ma98} have examined the effect of quantum
interference between the atomic transitions on the CPT and have demonstrated
that the CPT effect strongly depends on the cross-damping term $\Gamma _{12}$
and disappears when $\Gamma _{12}=\Gamma $.

The CPT effect and its dependence on quantum interference can be easily
explained by examining the population dynamics in terms of the superposition
states $|{s}\rangle $ and $|{a}\rangle $. We assume that a three-level $%
\Lambda $-type atom is composed of a single upper state $\left|
3\right\rangle $ and two ground states $\left| 1\right\rangle $ and $\left|
2\right\rangle $, and that the upper state is connected to the lower states
by transition dipole moments $\mbox{\boldmath $\mu$}_{31}$ and $%
\mbox{\boldmath $\mu$}_{32}$.

Introducing superposition operators $S_{s}^{+}=\left( S_{s}^{-}\right)
^{\dagger }=\left| 3\right\rangle \left\langle s\right| $ and $%
S_{a}^{+}=\left( S_{a}^{-}\right) ^{\dagger }=\left| 3\right\rangle
\left\langle a\right| $, where $\left| s\right\rangle $ and $\left|
a\right\rangle $ are the superposition states 
\begin{eqnarray}
|{s}\rangle  &=&\frac{1}{\sqrt{\Gamma _{1}+\Gamma _{2}}}\left( \sqrt{\Gamma
_{1}}|{1}\rangle +\sqrt{\Gamma _{2}}|{3}\rangle \right) ,  \nonumber \\
|{a}\rangle  &=&\frac{1}{\sqrt{\Gamma _{1}+\Gamma _{2}}}\left( \sqrt{\Gamma
_{2}}|{1}\rangle -\sqrt{\Gamma _{1}}|{3}\rangle \right) ,  \label{99}
\end{eqnarray}
then, in the basis of the superposition states~(\ref{99}) the Hamiltonian of
the system can be written as 
\begin{eqnarray}
H^{\prime } &=&-\hbar \left\{ \left( \Delta _{L}-\frac{1}{2}\Delta ^{\prime
}\right) S_{s}^{-}S_{s}^{+}+\left( \Delta _{L}+\frac{1}{2}\Delta ^{\prime
}\right) S_{a}^{-}S_{a}^{+}\right.   \nonumber \\
&&+\left. \Delta _{c}\left( S_{s}^{-}S_{a}^{+}+S_{a}^{-}S_{s}^{+}\right) +%
\frac{1}{2}\frac{\sqrt{\Gamma _{1}}\Omega }{\sqrt{\Gamma _{1}+\Gamma _{2}}}%
\left( S_{s}^{+}+S_{s}^{-}\right) \right\} ,  \label{100}
\end{eqnarray}
where 
\begin{equation}
\Delta ^{\prime }=\frac{1}{\Gamma _{1}+\Gamma _{2}}\left[ \left( \Gamma
_{1}-\Gamma _{2}\right) \Delta +4\delta _{12}\sqrt{\Gamma _{1}\Gamma _{2}}%
\right] .  \label{101}
\end{equation}
and 
\begin{equation}
\Delta _{c}=\frac{1}{\Gamma _{1}+\Gamma _{2}}\left[ \delta _{12}\left(
\Gamma _{1}-\Gamma _{2}\right) -\Delta \sqrt{\Gamma _{1}\Gamma _{2}}\right] .
\label{102}
\end{equation}
As before, $\Delta =\omega _{1}-\omega _{2}$, $\Delta _{L}=\omega _{L}-\frac{%
1}{2}\left( \omega _{1}+\omega _{2}\right) $, and we have assumed that $%
\Omega _{1}=\Omega _{2}=\Omega $.

From the master equation~(\ref{29}) with the Hamiltonian~(\ref{100}), we
derive the following equation of motion for the population $\rho_{aa}$ of
the antisymmetric state 
\begin{equation}
\dot{\rho}_{aa}=\frac{2\Gamma _{1}\Gamma _{2}}{\Gamma _{1}+\Gamma _{2}}%
\left( 1-p\right) \rho_{33}-i\Delta_{c}\left(\rho_{as}-\rho_{sa}\right) .
\label{103}
\end{equation}

The equation of motion~(\ref{103}) allows us to analyze the conditions for
population trapping in the driven $\Lambda $ system. In the steady-state $(%
\dot{\rho}_{aa}=0)$ with $p\neq 1$ and $\Delta _{c}=0$ the population in the
upper state is zero:  $\rho _{33}=0$. Thus the state $\left| 3\right\rangle $
is not populated despite the fact that it is continuously driven by the
laser. In this case the population is entirely trapped in the antisymmetric
superposition of the ground states. This is the coherent population trapping
effect. However, for $p=1$ and $\Delta _{c}=0$ the antisymmetrical state
decouples from the interactions, and then the steady-state population $\rho
_{33}$ is non-zero~\cite{java}. This shows that coherent population trapping
is possible only in the presence of spontaneous emission from the upper
state to the antisymmetric superposition state. Thus, we can conclude that
quantum interference has a destructive effect on coherent population
trapping. Menon and Agarwal~\cite{ma98} have shown that the CPT effect can
be preserved in the presence of quantum interference provided that the atom
is driven by two coherent fields, each coupled to only one of the atomic
transitions.

\begin{figure}
\begin{center}
\includegraphics[height=2.75in,width=4in]{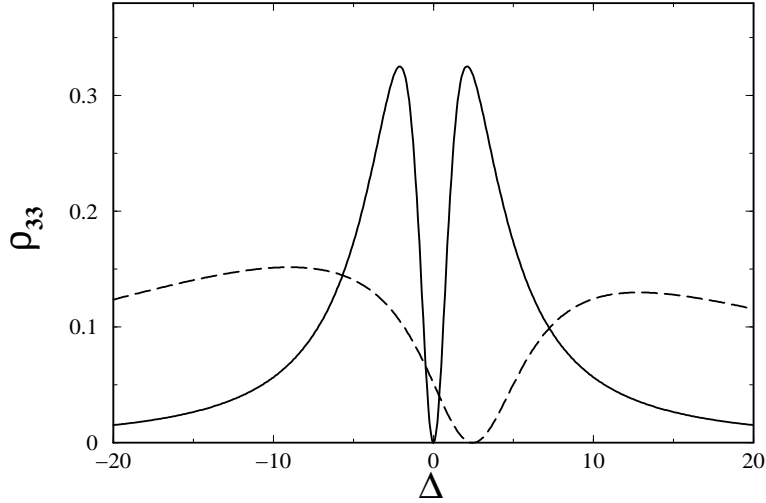}
\caption{The stationary population of the upper state
$\left|3\right\rangle$ of a $\Lambda$-type atom as a function of the splitting $\Delta$ for $\Delta_{L}=0, \Omega=5\Gamma_{1},
\delta_{12}=0.1\Gamma_{1}, p=0.5$ and different $\Gamma_{2}$:
$\Gamma_{2}=\Gamma_{1}$ (solid line), $\Gamma_{1}=50\Gamma_{2}$ (dashed line).}
\end{center}
\end{figure}

According to Eq.~(\ref{103}) the CPT can also be destroyed by the presence
of the coherent interaction $\Delta _{c}$ between the symmetric and
antisymmetric states. This is shown in Fig.~9, where we plot the
steady-state population $\rho _{33}$ as a function of $\Delta $ for
different values of $\Gamma _{2}/\Gamma _{1}$. It is evident that the
cancellation of the population $\rho _{33}$ appears only at $\Delta _{c}=0$,
i.e. in the absence of the coherent interaction between the antisymmetric
and symmetric states. For $\Gamma _{1}=\Gamma _{2}$ the cancellation appears
at $\Delta =0$, while for $\Gamma _{1}\neq \Gamma _{2}$ the effect shifts
towards non-zero $\Delta $ given by 
\begin{equation}
\Delta =\frac{\Gamma _{1}-\Gamma _{2}}{\sqrt{\Gamma _{1}\Gamma _{2}}}\delta
_{12}.  \label{104}
\end{equation}
The shift depends on the ratio $r$, and for either $r\ll 1$ or $r\gg 1$ can
be large despite $\delta _{12}$ being very small. Therefore, the vacuum
induced coherent coupling can be experimentally observed in the $\Lambda $
system as a shift of the zero of the population $\rho _{33}$ of the upper
state~\cite{afs}.

We have shown in Section~\ref{sec6a} that a laser field can drive the $V$%
-type system into the antisymmetric (trapping) state through the coherent
interaction between the symmetric and antisymmetric states. Recently, Akram~%
{\it et al.}~\cite{afs} have shown that in the $\Lambda$ system there are no
trapping states to which the population can be transferred by the laser
field. This can be illustrated by calculating the transition dipole moments
between the dressed states of the driven $\Lambda$ system. The procedure of
calculating the dressed states of the $\Lambda $ system is the same as for
the $V$ system. The only difference is that now the eigenstates of the
unperturbed Hamiltonian $H_{0}$ are $\left| 3,N-1\right\rangle ,\left|
1,N\right\rangle ,\left| 2,N\right\rangle $, and the dressed states are
given by 
\begin{eqnarray}
\left| +,N\right\rangle &=&\frac{1}{\sqrt{2}}\left[ -\alpha \left|
a,N\right\rangle +\left| s,N\right\rangle -\sqrt{2}\beta \left|
3,N-1\right\rangle \right] ,  \nonumber \\
\left| 0,N\right\rangle &=&-\sqrt{2}\beta \left| a,N\right\rangle +\alpha
\left| 3,N-1\right\rangle ,  \nonumber \\
\left| -,N\right\rangle &=&\frac{1}{\sqrt{2}}\left[ -\alpha \left|
a,N\right\rangle -\left| s,N\right\rangle -\sqrt{2}\beta \left|
3,N-1\right\rangle \right] .  \label{105}
\end{eqnarray}

Although the dressed states (\ref{105}) are similar to that of the $V$
system (see Eq.~(\ref{95})), there is a crucial difference between the
transition dipole moments. For the $\Lambda $ system the transition dipole
moments between the dressed states $\left| i,N+1\right\rangle $ and the
state $\left| 0,N\right\rangle $ of the manifold below are all zero, but
there are non-zero transition dipole moments between $\left|
0,N\right\rangle $ and the dressed states $\left| i,N-1\right\rangle $ of
the manifold below 
\begin{eqnarray}
\left\langle N,0\right| \mbox{\boldmath
$\mu$}\left| \pm ,N-1\right\rangle  &=&\pm \alpha \mbox{\boldmath $\mu$}, 
\nonumber \\
\left\langle N,0\right| \mbox{\boldmath
$\mu$}\left| 0,N-1\right\rangle  &=&0.  \label{106}
\end{eqnarray}
Therefore, population is unable to flow into the state $\left|
0,N\right\rangle $, but can flow away from it. If $\Delta =0$ then $\alpha =0
$, and the state $\left| 0,N\right\rangle $ completely decouples from the
remaining states. For $\Delta \neq 0$ the state $\left| 0,N\right\rangle $
is coupled to the remaining states, but does not participate in the dynamics
of the system because it cannot be populated by transitions from the other
states. Thus, there is no trapping state among the dressed states of the
driven $\Lambda $ system.

\section{Summary}

\label{sec8}

In this paper we have discussed quantum interference effects in optical
beams and radiation fields emitted from atomic systems. We have illustrated
the effects using the first- and second-order correlation functions of
optical fields and atomic dipole moments. We have explored the role of the
correlations between radiating systems and have presented examples of
practical methods to implement two systems with non-orthogonal dipole
moments. Moreover, we have derived general conditions for quantum
interference in a two-atom system and for control of spontaneous emission. We
have shown that the cancellation of spontaneous emission does not
necessarily lead to population trapping. The population can be trapped in a
dark state only if the state is completely decoupled from any interactions.
Finally, we have presented quantum dressed-atom models of cancellation of
spontaneous emission, amplification on dark transitions, fluorescence
quenching, and coherent population trapping.\\
\vspace*{0.15in}\\
\textbf{Acknowledgments}\\
This work was supported by the United Kingdom Engineering and Physical
Sciences Research Council and the Australian Research Council.
We wish to thank Helen Freedhoff, Paul Berman,
Peng Zhou, Terry Rudolph, Howard Wiseman, Uzma Akram, and Jin Wang for
many helpful discussions on quantum interference.

\end{document}